\documentclass[a4paper,fleqn,usenatbib]{mnras}
\usepackage{newtxtext,newtxmath}
\usepackage[T1]{fontenc}
\usepackage{ae,aecompl}
\usepackage{natbib,textcomp,placeins,epsfig,graphicx,color,amsmath,amssymb,upgreek,listings,epstopdf}

\newcommand{\logg}{$\log g$} 
\newcommand{\teff}{$T_{\mbox{\footnotesize eff}}$}
\newcommand{\al}{$\alpha$}
\newcommand{\mgas}{M_{\rm gas}} 
\newcommand{\mocc}{m_{\rm O}^{\rm cc}}
\newcommand{\mfecc}{m_{\rm Fe}^{\rm cc}} 
\newcommand{\mfeIa}{m_{\rm Fe}^{\rm Ia}} 
\newcommand{\mxcc}{m_{\rm X}^{\rm cc}} 
\newcommand{\mxIa}{m_{\rm X}^{\rm Ia}} 
\newcommand{\mdotstar}{\dot{M}_*} 
\newcommand{\mdotout}{\dot{M}_{\rm out}}
\newcommand{\mdotinf}{\dot{M}_{\rm inf}} 
\newcommand{\Zo}{Z_{\rm O}}
 
\newcommand{\ZfeIa}{Z^{\rm Ia}_{\rm Fe}} 
\newcommand{\Zfecc}{Z^{\rm cc}_{\rm Fe}} 
\newcommand{\Zoeq}{Z_{\rm O,eq}}

\newcommand{\feh}{[{\rm Fe}/{\rm H}]} 
\newcommand{\afe}{[\alpha/{\rm Fe}]}
\newcommand{\xfe}{[{\rm X}/{\rm Fe}]}
\newcommand{\ofe}{[{\rm O}/{\rm Fe}]} 
\newcommand{\ohh}{[{\rm O}/{\rm H}]}

\newcommand{\taustar}{\tau_*} \newcommand{\tausfh}{\tau_{\rm sfh}}
\newcommand{\taudep}{\tau_{\rm dep}} \newcommand{\taubar}{\bar{\tau}}

\newcommand{\Gyr}{\,{\rm Gyr}} 
 \newcommand{\kpc}{\,{\rm kpc}}

\newcommand{\tform}{t_{\rm form}}
\newcommand{\Rsol}{R_{\rm sol}}

\title[Age-resolved chemistry of giants]{Age-resolved chemistry of red giants in the solar neighbourhood}

\author[D. K. Feuillet et al.]{Diane K. Feuillet,$^{1}$\thanks{email: feuillet@mpia.de} 
Jo Bovy,$^{2,3}\thanks{Alfred P. Sloan Fellow}$ 
Jon Holtzman,$^{4}$ 
David H. Weinberg,$^{5}$ 
\newauthor
D. A. Garc\'ia-Hern\'andez,$^{6,7}$ 
Fred R. Hearty,$^{8}$
Steven R. Majewski,$^{9}$
\newauthor
Alexandre Roman-Lopes,$^{10}$
Jan Rybizki,$^{1}$
Olga Zamora$^{6,7}$
\\
$^{1}$Max-Planck-Institut f\"ur Astronomie, K\"onigstuhl 17, D-69117 Heidelberg, Germany\\
$^{2}$Department of Astronomy and Astrophysics, University of Toronto, 50 St. George Street, Toronto, ON, M5S 3H4, Canada\\
$^{3}$Dunlap Institute for Astronomy and Astrophysics, University of Toronto, 50 St. George Street, Toronto, Ontario, M5S 3H4, Canada\\
$^{4}$Department of Astronomy, New Mexico State University, Las Cruces, NM, 88003, USA\\
$^{5}$Department of Astronomy, The Ohio State University, Columbus, OH, 43210, USA\\
$^{6}$Instituto de Astrof\'isica de Canarias, E-38205 La Laguna, Tenerife, Spain\\
$^{7}$Universidad de La Laguna, Departmento de Astrof\'isica, E-38205 La Laguna, Tenerife, Spain \\
$^{8}$Department of Astronomy and Astrophysics, The Pennsylvania State University, University Park, PA 16802, USA \\
$^{9}$Department of Astronomy, University of Virginia, Charlottesville, VA 22904, USA \\
$^{10}$Departamento de F\'isica, Facultad de Ciencias, Universidad de La Serena, Cisternas 1200, La Serena, Chile
}

\date{Accepted XXX. Received YYY; in original form ZZZ}

\pubyear{2018}

\begin{document}
\label{firstpage}
\pagerange{\pageref{firstpage}--\pageref{lastpage}}
\maketitle

\begin{abstract}

In the age of high-resolution spectroscopic stellar surveys of the Milky Way,
the number of stars with detailed abundances of multiple elements is
rapidly increasing. These elemental abundances are directly
influenced by the evolutionary history of the Galaxy, but this can be difficult
to interpret without an absolute timeline of the abundance enrichment. We
present age-abundance trends for [M/H], [\al/M], and 17 individual elements
using a sample of 721 solar neighbourhood {\it Hipparcos} red giant stars
observed by APOGEE. These age trends are determined through a Bayesian
hierarchical modelling method presented by \citet{Feuillet2016}. We confirm that the [\al/M]-age
relation in the solar neighbourhood is steep and relatively narrow (0.20~dex age dispersion), as are the [O/M]- and [Mg/M]-age relations. The age trend of [C/N] is steep and smooth,
consistent with stellar evolution. The [M/H]-age relation has a mean age dispersion of 
0.28~dex and a complex overall structure. 
The oldest stars in our sample are those with the lowest and highest 
metallicities, while the youngest stars are those with solar metallicity. 
These results provide strong constraints on theoretical models of Galactic
chemical evolution (GCE). We compare them to the predictions of one-zone 
GCE models and multi-zone mixtures, both analytic and numerical.
These comparisons support the hypothesis that the solar neighbourhood 
is composed of stars born at a range of Galactocentric radii, and that the 
most metal-rich stars likely migrated from a region with earlier and more 
rapid star formation such as the inner Galaxy.
\end{abstract}


\section{Introduction}

Chemical abundances of stars can tell us a lot about the evolution of the Milky
Way and other spiral galaxies. The stellar populations of the Galactic disc,
where star formation has been occurring essentially continuously since its
formation, are particularly powerful tools to map the temporal evolution of
chemical elements. In order to get a clear picture of how the chemical
enrichment happened, we need samples of stars with detailed chemical abundances
from different nucleosynthesis channels and ages spanning the full
Galactic disc. High-resolution spectroscopic surveys like the first and second
phases of the Apache Point Observatory Galactic Evolution Experiment
\citep[APOGEE and APOGEE-2,][]{Majewski2017}, the Gaia-ESO Survey
(GES, \citealt{Gilmore2012}; \citealt*{Randich2013}), and the Galactic Archaeology with
HERMES survey \citep[GALAH,][]{Freeman2012, DeSilva2015, Martell2017} are providing
atmospheric parameters and measurements of multiple elemental abundances for
large samples of stars throughout the Galaxy. In conjunction, the Gaia mission
\citep{Gaia2016} is providing astrometric measurements for these same
stars.

While these surveys result in an unprecedented sample of stars with abundance
measurements across a huge range of distances with which the structure of stellar 
populations of different abundances can be mapped precisely \citep{Bovy2012a, 
Bovy2016a}, the complexity of Galactic chemical evolution (GCE) makes the interpretation 
of the global enrichment history
difficult. The time-scales of the chemical enrichment histories of stellar
populations are greatly affected by the local star formation history and gas
inflow/outflow (e.g. \citealt*{Chiappini1997}; \citealt{Dalcanton2007, Finlator2008}). It has also been shown that the migration of stars away from
their birth radius while conserving the orbit's eccentricity could be a
significant process in the Galaxy \citep[see][]{Wielen1996, Sellwood2002, Schonrich2009, Loebman2011}.
This process of radial migration would result in an observed chemical abundance
distribution at a given radius that has been altered from that of the native
population. The change in shape of the observed disc metallicity distribution
function (MDF), from negatively skewed in the inner Galaxy to positively skewed
in the outer Galaxy, provides further strong circumstantial evidence for this effect
\citep{Hayden2015, Loebman2016}.

Perhaps due to these complications from an evolutionary history that is not well known, recent studies have found that the traditionally discussed `thick' and `thin' Milky Way disk populations are not as separate in their spatial, kinematic, abundance, and age distributions as previously thought \citep[e.g.][]{Bensby2014, Hayden2017a}. Instead of the canonical `thick' and `thin' disk populations, in this work we occasionally refer to the high- and low-\al\, sequences, with the high-\al\, sequence equating more closely to the canonical `thick' disk and the low-\al\, sequence equating more closely to the canonical `thin' disk. These terms convey the directly observable quality of the two apparent sequences in the [\al/M] vs [M/H] distribution.\footnote{We discuss
the precise meaning of M below, but roughly speaking [M/H] refers to the
abundance of iron-peak elements and [\al/M] refers to the abundance of
$\alpha$-elements relative to iron-peak.}

In order to compare the stellar populations across the disc on a Galactic scale
and build a cohesive picture of Galactic evolution from chemical abundances and
astrometry, we need stellar age measurements to place all the stars on an
absolute timeline. However, stellar ages are difficult to determine for large
samples of field stars, such as those provided by the current high-resolution
spectroscopic surveys. While asteroseismology can provide precise ages
\citep[see][]{Chaplin2011}, it requires high time-cadence observations, and is
not currently possible for the full spectroscopic samples of hundreds of thousands stars.
Isochrone matching to spectroscopically derived atmospheric parameters can
provide reasonable age estimates for main sequence turn off and subgiant stars
(e.g. \citealt{Haywood2013}; \citealt*{Bensby2014}), but the age uncertainties become large ($\sim0.25$~dex)
for red giant stars if only spectroscopic parameters are known \citep[see
][hereafter F16]{Feuillet2016}. Although more precise ages can be determined
for subgiants, this is a relatively brief stage of stellar evolution, and the
higher luminosity of red giants makes them better suited to probing the full
Galactic disc and bulge. As a result, samples of fields stars with age
determinations have been fairly small and have not extended far beyond $\sim1$~kpc of the Sun, or the {\it Kepler} and CoRoT fields \citep[e.g.][]{Chaplin2011, Anders2017}. 
The exceptions to this limitation are the recent studies that have pushed out 
farther into the Galactic disc using C and N abundance 
ratios as a proxy for mass/age in red giants \citep{Masseron2015, Martig2016, Ness2016b}. This CN age 
method has been used to map the stellar populations structure in age and 
abundance for a large sample of stars \citep{Mackereth2017}. However, this 
method does require calibration either to models or empirically to samples with precise ages, as well as a correction for potential Galactic gradients in C and N. Therefore techniques directly connected 
to stellar physics that do not rely on elemental abundances are preferable. Fortunately, the age estimate of
field stars, particularly subgiants and giants, can be greatly improved if the
distance, and therefore the luminosity, is known.  Soon the number of
stars for which ages can be determined will greatly increase with the
availability of parallax measurements from Gaia \citep{Gaia2016}.

Much work has been done trying to determine whether an age-metallicity relation
exists for stars in the solar neighbourhood \citep[e.g.][]{Edvardsson1993,
Nordstrom2004, Bergemann2014}, and recent studies have found \al-element
enhancement, i.e. [\al/M], to have a tighter correlation with age than
metallicity, [M/H] \citep{Haywood2013, Bensby2014}. From these previous observational studies as well as studies of GCE models \citep[e.g.][]{Chiappini1997, Minchev2013}, it appears that the [\al/M]--[M/H]--age or kinematics--age space is a better diagnostic for the evolution of the high- and low-\al\, sequences. Using large samples of stars with CN ages or asteroseismology, recent work has found that the mean age of the high-\al\, sequence is distinctly older than the mean age of the low-\al\, sequence, and that there is little age evolution with increasing [M/H] and decreasing [\al/M] along the high-\al\, sequence \citep[][with age uncertainties of 0.2, 0.2, and 0.1~dex, respectively]{Ness2016b, Ho2017, Wu2017}. This suggests that the high-\al\, sequence formed rapidly, and almost entirely before the low-\al\, sequence. 

Until very recently, few
studies have looked at the relationship between age and individual elemental
abundance for more than one or two elements. \citet{Spina2016}, \citet{Adibekyan2016}, and \citet{Bedell2018} 
have explored trends of age with individual abundances of
solar-type stars, inspired by the strong, linear correlation between age and
the [Y/Mg] abundance ratio of solar twins found by \citet{DaSilva2012} and
\citet{Nissen2015} \citep[see also][]{Feltzing2017, Slumstrup2017}. These studies analysed 
small samples of dwarf stars (a few 100 at most) and
highlighted the discovery that certain abundance indicators can be used as
chemical clocks \citep[see also][]{VanEck2015}. While these tight age-abundance relations are extremely
interesting and potentially very useful, they have not been tested beyond the
solar neighbourhood where conditions of star formation were likely different.
Here we focus on a larger sample of red giant stars ($\sim 700$), and the aim of the present
work is not primarily to find an abundance ratio that can be used as a proxy
for age, but rather to characterise the enrichment of the individual elements
with time for the purposes of constraining the parameters of Galactic evolution
such as radial migration, gas inflow, and gas mixing in the disc, as well as nucleosynthetic yields. 

\citet{Jia2018} use the sample presented in F16 combined with the APOGEE Data Release~13 (DR13) elemental abundances to examine the age-abundance trends for Fe, Si, K, Ca, Mn, and Ni. They use the ages derived from the spectrophotometric mass and the red giant mass-age relation, which have uncertainties of 0.38~dex in log(age). \citet{Jia2018} find a large spread in abundance at all but the youngest ages, and that all but one of the most enhanced stars in [Si/M] and [Ca/M] are old. Smaller age uncertainties are possible for this sample using a Bayesian isochrone matching approach as discussed below and in F16.

In this paper, we use the sample of solar neighbourhood red giant stars observed
using the New Mexico State University (NMSU) 1-m~+~APOGEE set-up presented by F16 to examine the trends of
individual elemental abundance with age in the solar neighbourhood. This work is improving the analysis by using updated spectroscopic parameters and modeling the abundance-age trends as a function of each individual elements, instead of just [\al/M]. We find an interesting
turn over in the age-metallicity relation at high metallicities. The trends of
most of the individual \al-elements are consistent with the overall [\al/M]
trend. 
The hierarchical modeling method is briefly outlined in Section~\ref{sec:method}, and
fully described in F16. In Section~\ref{sec:results}, we present the age-abundance
trends for [M/H], [\al/M], and 17 individual elements. In Section~\ref{sec:GCE}, we present some GCE
models for qualitative interpretation of these observations. We summarise our
findings in Section~\ref{sec:conclusion}.

\section{Method} \label{sec:method}

\begin{figure}
\centering 
\includegraphics[clip, trim=2cm 0.5cm 0.5cm 1cm, width=0.49\textwidth]{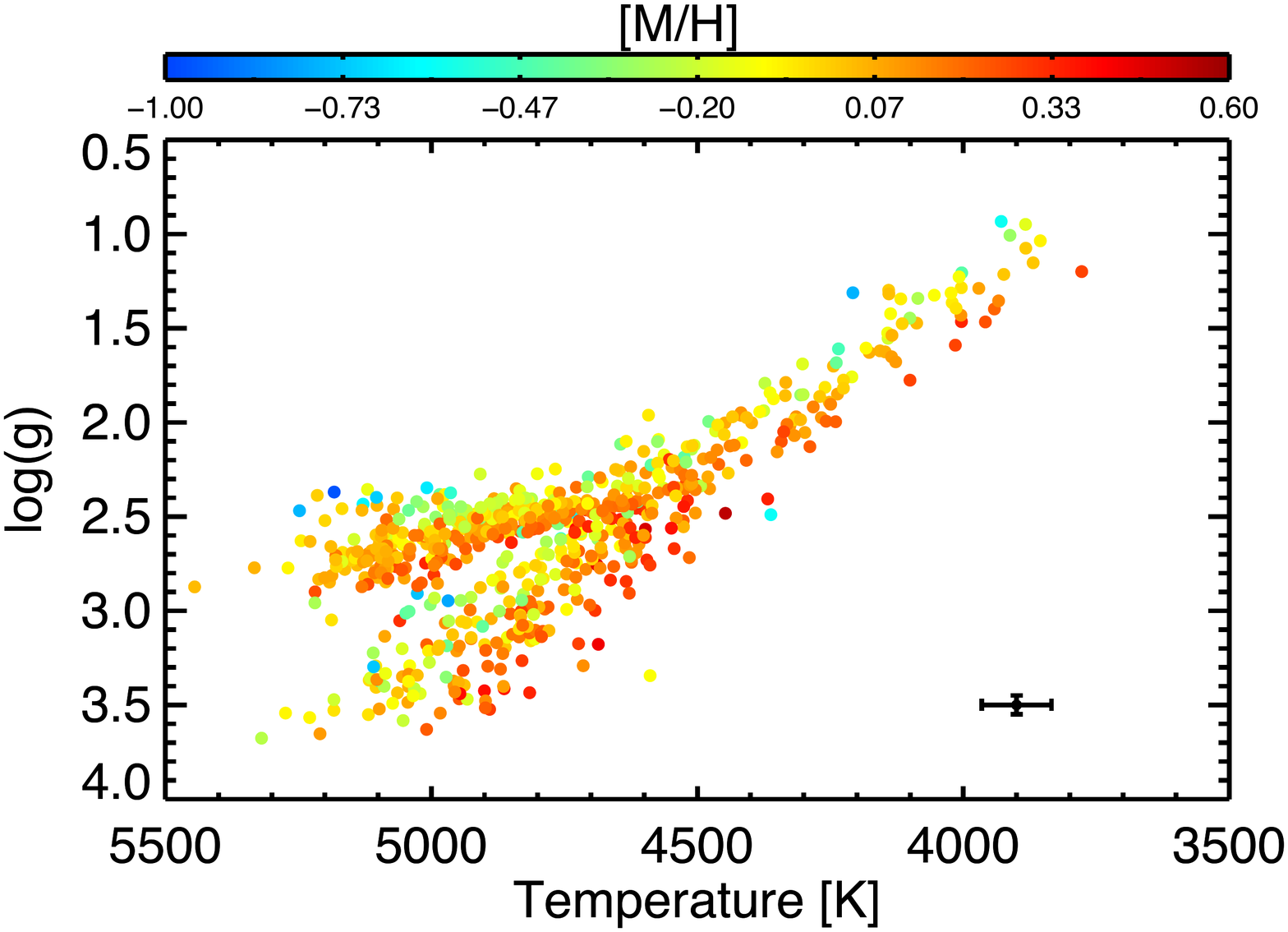} 
\caption{The spectroscopic Hertzsprung--Russell Diagram of the sample with DR14 parameters. The colour indicates the metallicity. The typical uncertainty is shown in the bottom right corner.} 
\label{fig:HRD} 
\end{figure}

\subsection{Observed sample}

We use a sample of 721 local red giant stars observed with the NMSU 1-m~+~APOGEE
capability presented in F16. These stars were selected to have $(J-K)_0 > 0.5$,
an absolute 2MASS $H$ magnitude brighter than 2, and a Hipparcos parallax error
less than 10~per cent \citep{VanLeeuwen2007}. The distance measurement is critical because, as was shown in
F16, using the absolute magnitude in addition to the spectroscopic parameters
to determine the age of giants reduces age uncertainties by $\sim 0.07$ dex.

In this work we use APOGEE calibrated atmospheric parameters and chemical
abundances from the Sloan Digital Sky Survey IV \citep[SDSS-IV,][]{Blanton2017, Gunn2006} Data Release 14
\citep[DR14,][]{Abolfathi2017}. The APOGEE target selection is described in 
\citet{Zasowski2013} and data reduction is described in \citet{Nidever2015}. 
The atmospheric parameters and chemical abundances are determined using the APOGEE Stellar Parameters
and Chemical Abundances Pipeline \citep[ASPCAP,][]{GarciaPerez2016}. ASPCAP
uses the code FERRE \citep{AllendePrieto2006} in combination with grids of
normalised stellar synthetic spectra \citep[see][]{Zamora2015}, as created with extensive atomic and molecular linelists \citep{Shetrone2015}, to simultaneously determine
\teff, \logg, [M/H], [\al/M], [C/M], [N/M], and
microturbulance through $\chi^2$ minimization. During this process, ASPCAP
fits for the overall abundance of metals [M/H] and the relative abundance of
\al-elements [\al/M], using O, Mg, Si, S, Ca, and Ti; the relative abundances
of non-\al\ elements (except for C and N) are set to solar ratios when
inferring [M/H]. The measurements of both [M/H] and [\al/M] are dominated by different elements in
different temperature regimes. The abundances of the individual elements are
then determined by fitting specific spectral windows containing relevant
absorption lines while varying the corresponding family of elements \citep[][2018, in preparation]{Holtzman2015}.  The DR14 abundances have been internally calibrated to remove \teff\, correlations in open clusters, similar to the corrections to DR12 described in \citet{Holtzman2015}; see Holtzman (2018, in preparation) for a description of the DR14 calibrations.
The APOGEE team has found that the metallicity parameter [M/H] closely correlates
with the [Fe/H] measurement \citep[][2018, in preparation]{Holtzman2015}. In this paper we will
assume that the ASPCAP measurements of [M/H] and [\al/M] can be taken as
proxies from [Fe/H] and [\al/Fe], and therefore compared to literature results
and theoretical models for these ratios.

In Fig.~\ref{fig:HRD} we display the spectroscopic Hertzsprung--Russell (HR) diagram colored by [M/H]
for our sample using DR14 calibrated parameters. In F16, we applied separate
calibrations of the surface gravities (\logg) for red clump (RC) stars and first-ascent red giants but,
this additional correction has been included in the internal APOGEE DR14 calibrations. 
Through comparisons to asteroseismic \logg\, measurements, it has been found that the \logg\, of massive secondary clump stars (mass $> 1.5 M_{\odot}$) has been overcorrected in DR14 by up to 0.2~dex and the true \logg\, is higher. While we do not have seismic information about the evolutionary stage of our stars, we use the mass calculations from F16 (with uncertainties of 0.11~dex, $\sim30\%$) and the position in \teff\, vs \logg\, space to determine that our sample has roughly 150 stars that could be affected by this \logg\, overcorrection. A higher \logg\, in the secondary clump stars would cause the age estimation from Bayesian isochrone matching to be younger. Therefore our age estimates for the massive secondary clump stars could be too old. We discuss below how this might affect each of our observed abundance-age trends in \S \ref{sec:results}.
We refer the reader to Holtzman (2018, in preparation) for a full discussion of the DR14 analysis and
calibrations. 

We use updated parallax measurements from the {\it
Tycho--Gaia} Astrometric Solution catalogue \citep[TGAS,][]{Lindegren2016, Brown2017} for those 
stars in TGAS with smaller uncertainties than the Hipparcos measurement. 
Approximately 60~per cent of the stars in our sample are included
in TGAS, the remaining stars are too bright for this catalogue. If the TGAS
parallax uncertainty is smaller than the Hipparcos uncertainty from
\citet{Anderson2012}, then we use the TGAS parallax and uncertainty. There are
now approximately 150 additional giant stars in APOGEE with TGAS parallax
uncertainties below 10~per cent that were not included in the F16 sample, but there is a dependence 
in the TGAS parallax uncertainty on \logg, because lower \logg\, stars are typically more 
distant and therefore have larger relative TGAS parallax uncertainties. To avoid biasing
our sample due to our parallax uncertainty cut, we do not include these
additional stars in our analysis. For future analyses of larger samples of
APOGEE-{\it Gaia} stars, a full characterisation of this \logg\, bias will be
needed.

\begin{figure*}
\centering 
\includegraphics[clip, trim=5cm 2.5cm 1cm 2.8cm, width=0.45\textwidth]{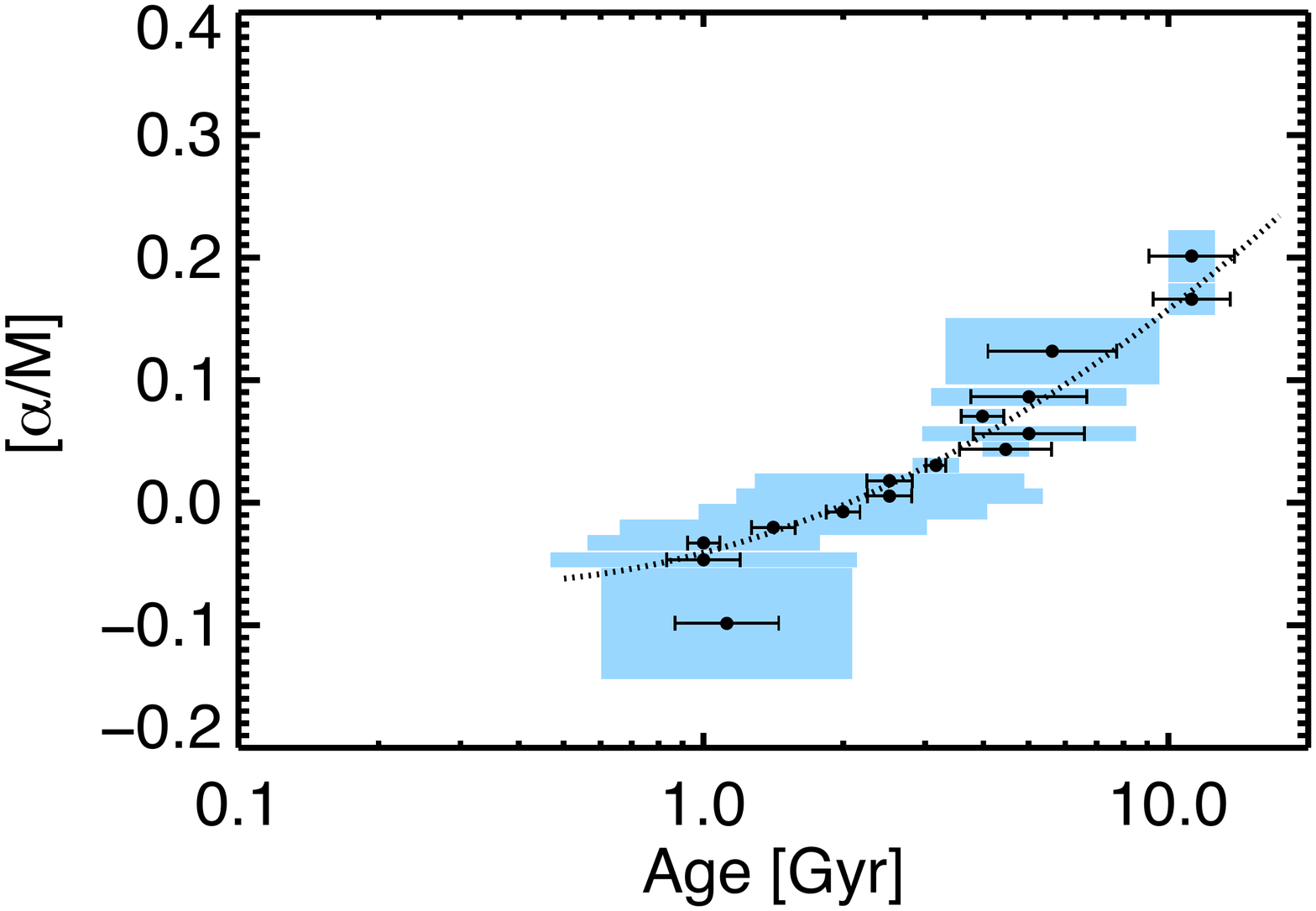} 
\includegraphics[clip, trim=1.5cm 0.5cm 0.5cm 1cm, width=0.49 \textwidth]{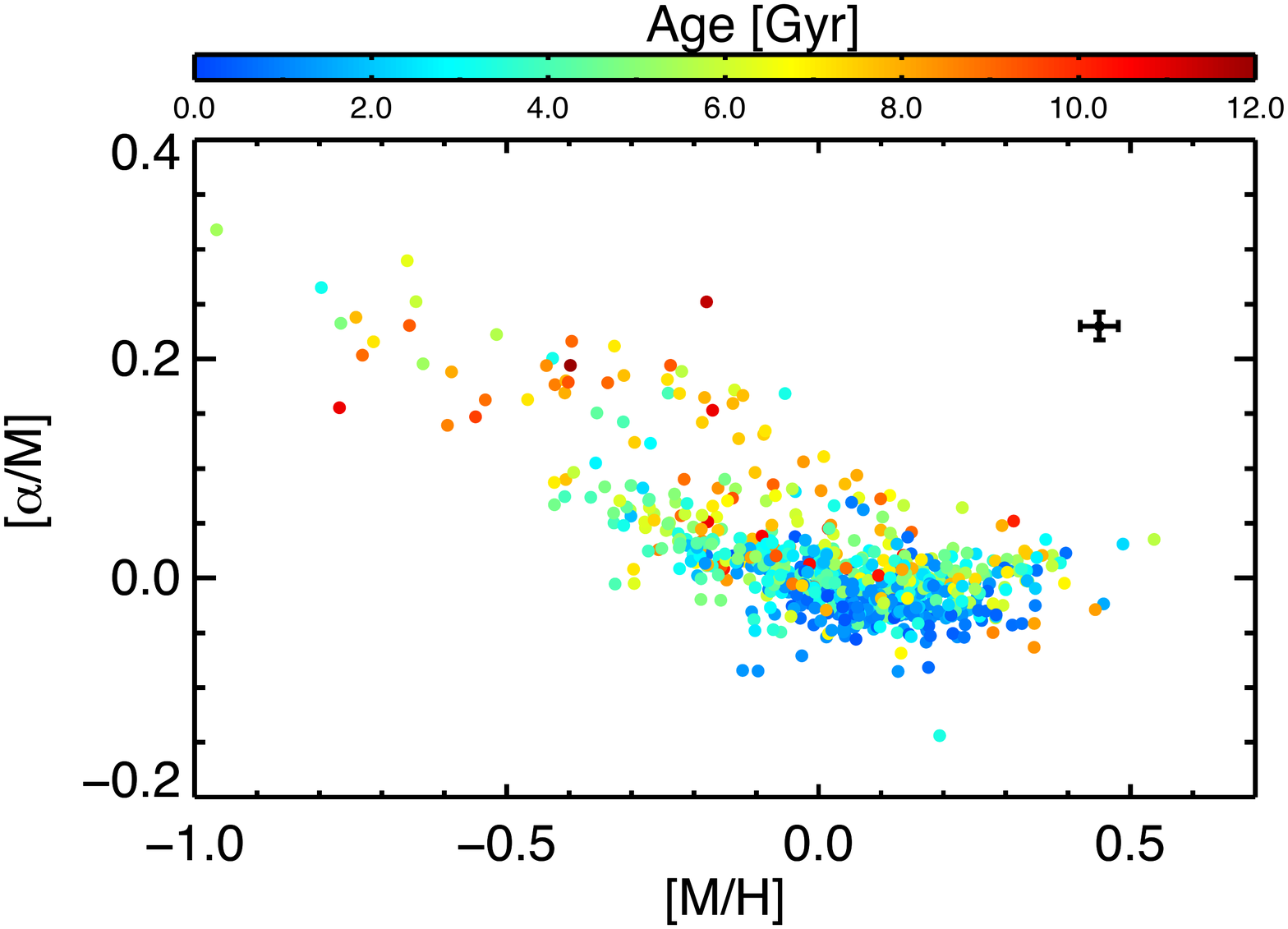}
\caption{Left: The parameters of a Gaussian+uniform model SFH determined as a
function of [\al/M]. The points indicate the mean age of the SFH and the mean
bin [\al/M]. The error bars indicate the uncertainty in the mean age
parameter. The shaded region indicates the age dispersion of the model SFH and
abundance range of the bin. We indicate a quadratic fit to the [\al/M]-log(age)
relation used for comparison to other elements with the dotted line, see
Equation \ref{eq:alpha_fit}. Right: The [\al/M] vs [M/H] distribution of the
sample using DR14 parameters, colored by the simple Bayesian age before hierarchical modeling is done. A typical
error bar is shown in the top right corner.} 
\label{fig:age_al} 
\end{figure*}

\subsection{Age determination}

We first redo the Bayesian isochrone matching analysis using the DR14 parameters.
Following the method described in F16, we calculate the age
probability distribution function (PDF) for each star using a simple Bayesian
isochrone matching to effective temperature (\teff),
surface gravity (\logg), [M/H], and absolute Tycho $V$-band magnitude
($M_{V_{T}}$). We assume a Chabrier lognormal initial mass function
\citep[IMF,][]{Chabrier2001}, a metallicity distribution function that is flat
within the uncertainties of [M/H], and a flat star formation history (SFH) in
log(age), as dictated by the isochrone age grid.

The full age PDFs for each star using DR14 parameters are available online through the public data repository Zenodo
(DOI:10.5281/zenodo.1199541).
As in F16, these age PDFs are often highly non-Gaussian with
multiple peaks or a significant tail to large or small ages. A single age from
the simple Bayesian isochrone matching is determined to be the probability
weighted mean of the age PDF. The right panel of
Fig.~\ref{fig:age_al} shows the [\al/M] vs [M/H] figure colored by this Bayesian
age. We show these ages to give a sense of the age distribution with [\al/M]
and [M/H] independent of the hierarchical modeling, but we note that the median
uncertainty on these Bayesian ages is 0.16~dex in log(age) or $\sim2$~Gyr at an
age of 5~Gyr. 

To produce more robust age-abundance trends than is possible using the ages of individual 
stars determined as the mean of the Bayesian age PDF, we use the hierarchical modeling
technique presented in F16. This method uses the full age PDFs produced by a simple
Bayesian isochrone matching to constrain the parameters of a model SFH. By
using the combined age PDFs of mono-abundance stars we are able to preserve the
information in the individual non-Gaussian PDFs, which is lost when a single
mean age is taken instead. As in F16, we use a Gaussian + uniform model SFH to
determine the mean age and age dispersion of the mono-abundance sample.  These parameters 
are determined to be the most likely mean and dispersion of the Gaussian function 
and are not strongly influenced by any outlier stars. For a mono-abundance sample in which most of the age PDFs peak at very different ages, the dispersion of the Gaussian will be large. By contrast, a mono-abundance sample in which most of the age PDFs peak at a similar age and only a few PDFs peak at a different age, the dispersion will be small and the presence of the outlier stars will not be obvious from the SFH parameters. 
This model is a Gaussian in log(age) with a prior probability of 7.5~per cent that a given star is drawn from a uniform distribution in log(age) instead of the Gaussian
distribution, allowing for outlier stars. The choice of 7.5~per cent for the outlier
fraction is arbitrary, but we find the exact value does not have a large effect
on the Gaussian parameters. The outlier fraction is more important when using the modeled SFH to find individual ages of stars. An outlier fraction of 15~per cent produces the same SFH parameters as 7.5~per cent.
In F16, the SFH model was fit to subsamples of stars with a
single [\al/M], but here we also fit the model SFH to subsamples in bins of
[M/H] and the individual [X/M] abundance ratios. An independent hierarchical model is 
created for each element, resulting in different SFH models. 

To create the abundance bins for each element the stars are sorted by the given abundance, then starting at one abundance 
edge a bin is created with some minimum width in abundance. However, if fewer than 15 stars are 
contained within the minimum bin width, the bin is expanded until it contains 15 stars. Here, the
minimum bin width is set to be the mean uncertainty in the abundances of each
element. We tested using different bin widths and minimum numbers of stars and
find that the results produced similar trends regardless of binning. We also found
the results were robust against binning from low to high abundance versus high to
low abundance. We present results binned from low to high abundance, therefore
the lowest abundance bin can be strongly influenced by stars with anomalously
low abundances, and the highest abundance stars are sometimes not included due
to a lack of stars. It should be noted that full systematic uncertainties due
to spectroscopic methods and stellar models are not included in our analysis.
For a full description of the hierarchical modeling analysis techniques, we
refer the reader to F16.

\section{Age trends} \label{sec:results}
\subsection{[\al/M] and overall metallicity}

In F16, we found that the mean age of stars steeply increases with [\al/M].
This result is reproduced here in Fig.~\ref{fig:age_al} using DR14 [\al/M]
values. In this figure, the points indicate the mean age of the model SFH as a
function of the [\al/M] bin. The error bars
indicate the uncertainty in the mean age parameter. This uncertainty is given
as the variance in the probability distribution function of the mean age
parameter. The blue shaded regions indicate the value of the age dispersion
parameter for the model SFH and the range of the abundance bin. All figures
with the mean hierarchical age on the $x$-axis are shown as linear age with a
log scale. This choice is to allow for easier comparisons with other studies while
retaining the scale used in the hierarchical modeling method. 

The resulting [\al/M] trend using DR14 values is qualitatively the same as in
F16 using DR12 values, showing a steep increase in mean age with [\al/M], even
at young ages. Recent studies have found a steepening of the age-[\al/M]
relation at older ages \citep[see e.g.][]{Haywood2013,Bergemann2014}, which is
in agreement with our results. We also find that the lower [\al/M] bins have 
larger age dispersions, indicating a larger range of ages
represented by lower [\al/M] stars, which is consistent with previous studies (e.g. \citealt*{Ramirez2013}; \citealt{Haywood2013, Bensby2014, Hayden2017a}). The dotted
line in Fig.~\ref{fig:age_al} indicates a fit to the [\al/M]-age relation used to
compare the [\al/M] trend with other elements in the figures below. The
equation used is quadratic in log(age) [yr], $\tau$: 
\begin{equation}
\label{eq:alpha_fit} 
	[\alpha/M] = 7.0589 - 1.6775 \tau + 0.0987 \tau^2 ~.
\end{equation} 
The right panel of Fig.~\ref{fig:age_al} shows the [\al/M] vs
[M/H] distribution of this sample, colored by the single age determined as
the mean of the simple Bayesian isochrone matching. The age used here is the mean of the Bayesian age PDF, and is not yet informed by the hierarchical modeling. The typical [\al/M] and
[M/H] uncertainty is indicated in the top right corner. The typical age
uncertainty in the Bayesian age is 0.16~dex. The trend between [\al/M] and age
can be recovered by eye from this figure, but due to the large age
uncertainties and non-Gaussian Bayesian age PDFs, it is much more clear when
the hierarchical modeling is done.

\begin{figure*}
\centering 
\includegraphics[clip, trim=5cm 2.5cm 1cm 2.8cm, width=0.49\textwidth]{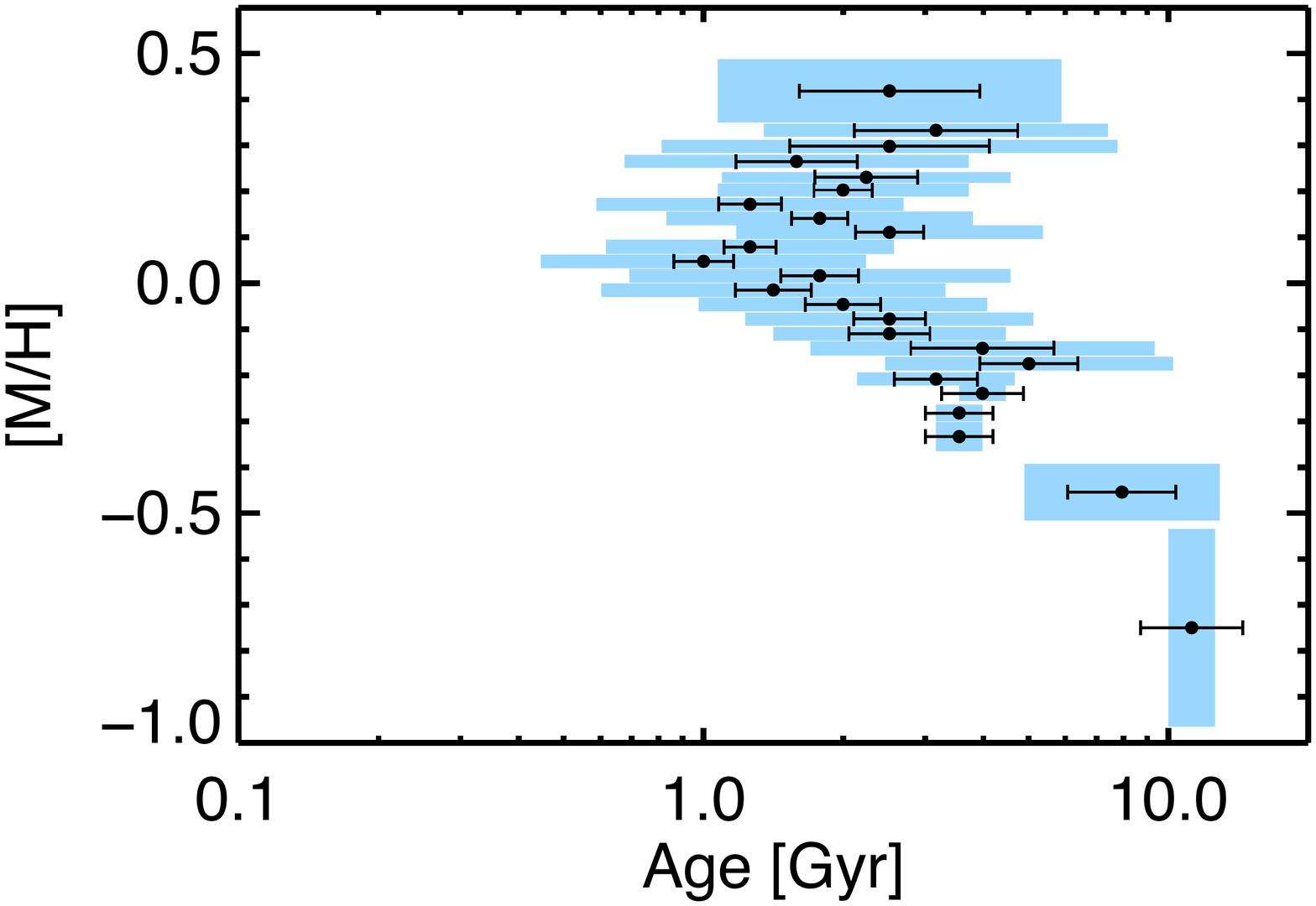} 
\includegraphics[clip, trim=5cm 2.5cm 1cm 2.8cm, width=0.49\textwidth]{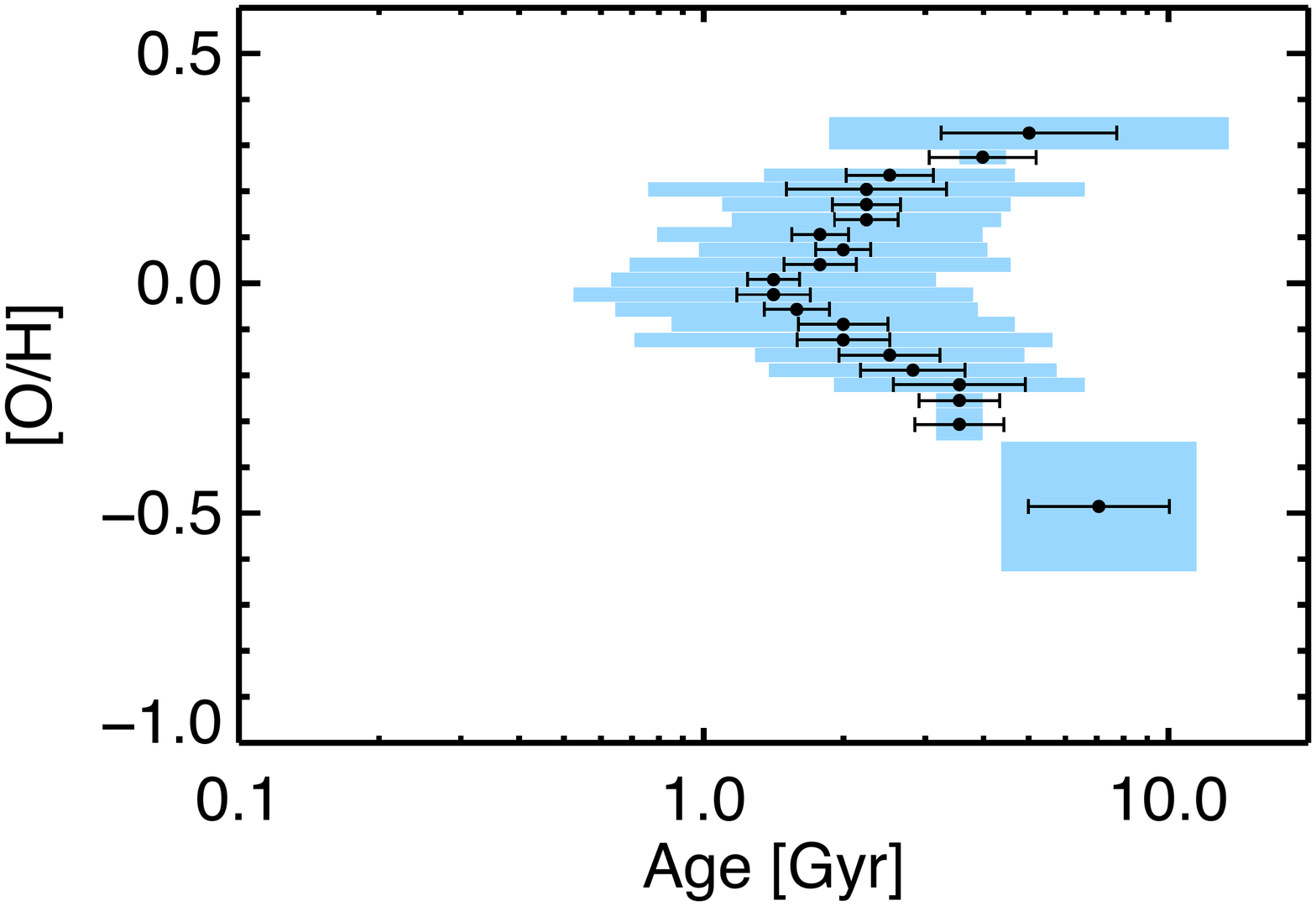}
\caption{The same as left panel of Fig.~\ref{fig:age_al} but for [M/H] and [O/H]. In
[M/H] and [O/H], both the lowest abundance and highest abundance stars are
older than the solar abundance stars.} 
\label{fig:age_fe_o} 
\end{figure*}

\begin{figure*}
\centering 
\includegraphics[angle=90, trim=1cm 1cm 1cm 1cm, width=1.0\textwidth]{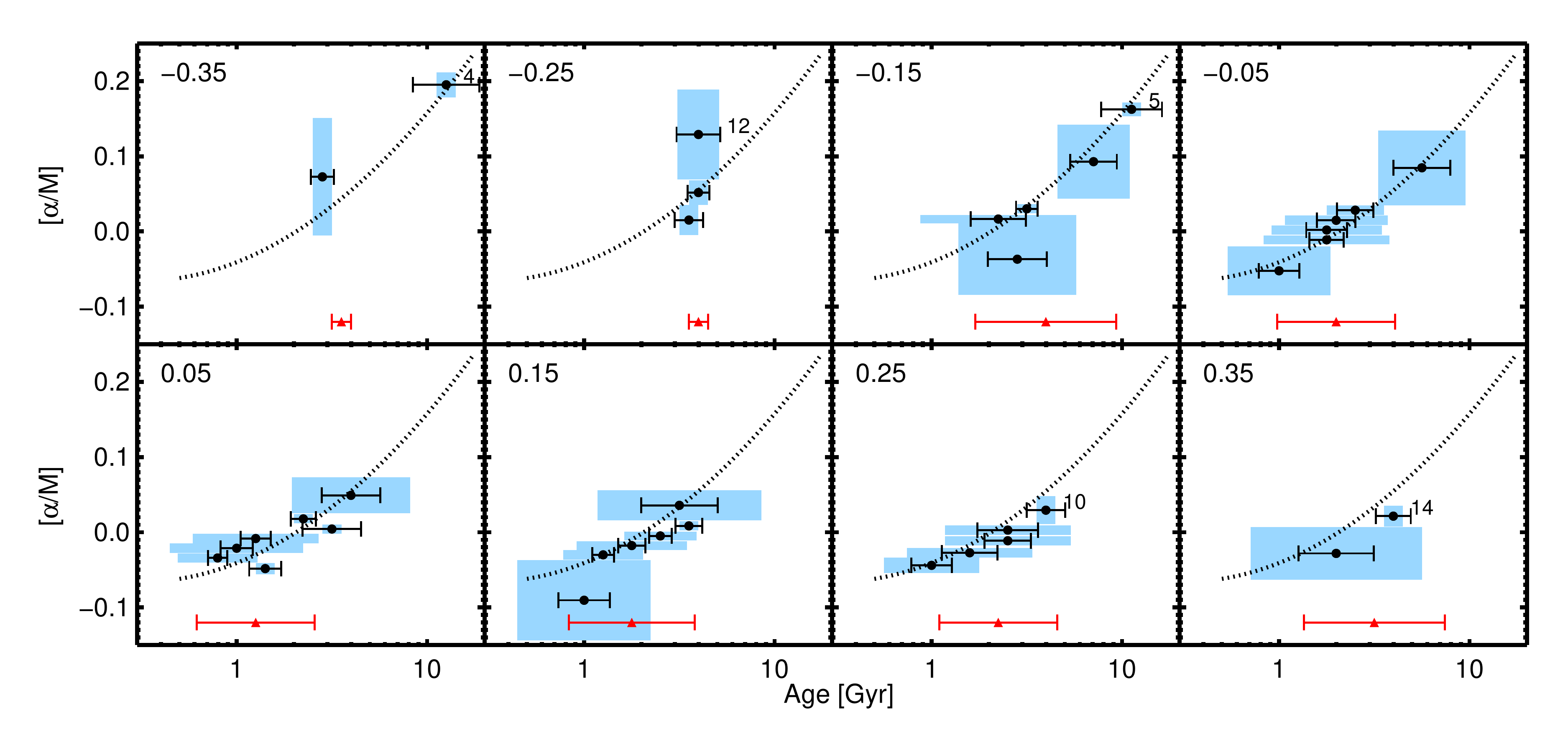} 
\caption{Hierarchical modeling of the SFH as a function of both [M/H] and [\al/M]. Each panel gives
the [\al/M]-age relation within a single [M/H] bin. The point gives the mean
age and the error bars indicate the uncertainty in the mean. The blue shaded region indicates 
the age dispersion and the abundance width. For points with fewer than
15 stars, the number of stars used is given next to the point. The [M/H] bins
are 0.1 dex wide and the central value is shown in the upper left corner of the
panel. The fit to the overall [\al/M]-age relation in Fig.~\ref{fig:age_al} and
Equation~\ref{eq:alpha_fit} is shown by the dotted line. In the bottom of each
panel in red is given the mean age and age dispersion of the [M/H] bin from
Fig.~\ref{fig:age_fe_o} that is closest to the panel central metallicity. The
binning in metallicity was done differently in these two figures, therefore
there is not a directly corresponding bin for each panel.} 
\label{fig:age_alpha_fe}
\end{figure*}

Fig.~\ref{fig:age_fe_o} shows the results of a model SFH constrained as a
function of [M/H] in the left panel and [O/H] in the right panel. [M/H] is a more
complicated tracer of Galactic chemical evolution than the \al-elements due to
the contributions to [M/H] from both core-collapse supernovae and type Ia
supernovae. Therefore, we also examine [O/H] as a tracer of the overall
metallicity. The age trends with [M/H] and [O/H] are similar. 
We find that the mean age is oldest for the most metal-poor stars in our sample and
decreases with increasing metallicity until solar metallicity. We
see a turn over in the age-metallicity relation above solar metallicity so that
the most metal-rich stars have a mean age that is older than the solar
metallicity stars. Fig.~\ref{fig:age_al} shows the [\al/M] vs [M/H] distribution
for the sample. The distribution of the low-\al\, sequence shows an upturn in
[\al/M] above solar metallicity. The most metal-rich stars actually have solar
[\al/M], and therefore have an age consistent with the steep and smooth age
increase with [\al/M]. This observation that the most metal-rich stars are not
the youngest stars is consistent with observations from the Geneva-Copenhagen
Survey \citep[see fig.~16 of][]{Casagrande2011}, which find the metal-rich
end of the solar neighbourhood metallicity distribution function is dominated by
intermediate-age and old stars. This is also consistent with fig.~17 of F16, which shows that the younger
stars have a smaller spread in metallicity and that the most metal-rich stars
actually have intermediate ages of 1--4~Gyr.

In Fig.~\ref{fig:age_fe_o} we see a large age dispersion at all metallicities, suggesting a broadly peaked age distribution. 
This is also observed by other studies of the solar neighbourhood, which find a large spread in
metallicity at all ages \citep[see][]{Edvardsson1993, Bensby2014, Bergemann2014, Mackereth2017}. The results of \citet{Ness2016b}, \citet{Ho2017}, and \citet{Wu2017} suggest that this broad age distribution at a given [M/H] should be double-peaked and a function of [\al/M] due to the presence of the older, high-\al\, sequence and the younger low-\al\, sequence. 

We further investigate the [\al/M]--[M/H]--age
space by performing a hierarchical modeling on stars with similar [\al/M] and
[M/H] abundances. To do this, we first bin the stars by [M/H], using bins with
a constant width of 0.1~dex. Within each [M/H] bin, we bin in [\al/M] using at
least 15 stars in each bin and a minimum bin width equal to the mean ASPCAP
uncertainty in [\al/M], in the same way as is done in Fig.~\ref{fig:age_al}. The
results of this exercise are shown in Fig.~\ref{fig:age_alpha_fe}. Each panel
shows the [\al/M]--age relation for a single [M/H] bin. The bin central [M/H]
is given in the upper left corner of each panel. The points indicate the mean
age and the error bars indicate the uncertainty in the mean. 
The blue shaded region 
shows the age dispersion and the abundance width. Due to the small number of
high [\al/M] stars in our sample, the highest [\al/M] bins do not all contain
15 stars, but we show these bins in order to get a sense of the age
distribution across the sample. For bins with fewer than 15 stars, the number
of stars is indicated next to the point in Fig.~\ref{fig:age_alpha_fe}. The fit
to the overall [\al/M]--age relation, see Equation~\ref{eq:alpha_fit}, is shown by
the dotted line. At the bottom of each panel there is a point in red. This is the
mean age and age dispersion of the overall [M/H]--age relation bin from 
Fig.~\ref{fig:age_fe_o} that is closest to the bin central metallicity. Because the
binning in metallicity was done differently in the two figures, there is not
always a corresponding [M/H] bin between Fig.~\ref{fig:age_fe_o} and~\ref{fig:age_alpha_fe}. 

It is clear from this figure that the large spread in the age metallicity
relation seen in Fig.~\ref{fig:age_fe_o} is due to a spread in [\al/M] at each
metallicity, and that, on average, the high [\al/M] stars are older than the low [\al/M] at
the same [M/H]. In Fig.~\ref{fig:age_alpha_fe} it can be seen that at any given
metallicity there is a range in both [\al/M] and age, which generally follows
the same [\al/M]--age relation that is found using the full sample. There could be outliers to this general trend in our sample, such as the young \al-rich stars noted by \citet{Martig2015} and \citet{Chiappini2015}. In F16, we noted the presence of some young [\al/M]-rich stars. Using the hierarchical modeling method, we obtain only the mean age and age distribution of the model SFH and not ages of individual stars. In our sample we find a small age dispersion in the highest [\al/M] bins suggesting there are very few intermediate or young stars at these abundances. However, the bins lower that [\al/M]=0.15 have larger age dispersions, which could be caused by the presence of young [\al/M]-rich stars.

Separating the metal-rich stars into high- and low-\al\ subsets does not
reveal any truly young population at the highest metallicities that could have been dominated by older, high-[\al/M] stars. In
Fig.~\ref{fig:age_alpha_fe}, as in Fig.~\ref{fig:age_fe_o}, the youngest stars have solar metallicity. The most
metal-rich stars are unlikely to be a continuation of simple star formation in
the solar neighbourhood as might be expected from their metallicity alone due to
their older age and slightly higher [\al/M] than the solar metallicity stars.
In addition, observations of young B--stars in the solar neighbourhood, which 
reflect the present-day interstellar medium, reveal solar metallicities 
\citep[e.g.][]{Nieva2012}. This suggests that the metal-rich stars must have
formed elsewhere. These stars could have formed outside of the solar
neighbourhood, perhaps the inner Galaxy where the mean [M/H] of stars is observed to be higher, 
and radially migrated to the solar radius.  
This is consistent with the behavior 
of the radial `donut' profile found by \citet{Bovy2016a} and \citet{Mackereth2017} 
using APOGEE observations of the disc. In these studies, the density profiles of the 
most metal-rich stars peaks in the inner Galaxy, suggesting that most metal-rich stars were formed inside the solar neighbourhood. 
In the full chemo-dynamical model of the Milky Way presented by \citet*{Minchev2013},
the metal-rich end of the solar neighbourhood
metallicity distribution function is composed primarily of stars born at a
galactocentric radius of 3--5 kpc, see their fig.~3. In the model, stars born at this radius have
primarily intermediate to old ages. We discuss comparisons of our observed
age--metallicity trend to multi-zone analytic chemical evolution models in
Section~\ref{sec:WAF}.

\begin{figure*}
\centering 
\includegraphics[clip, angle=90, trim=0.4cm 0.8cm 0.9cm 0.8cm, width=0.99\textwidth]{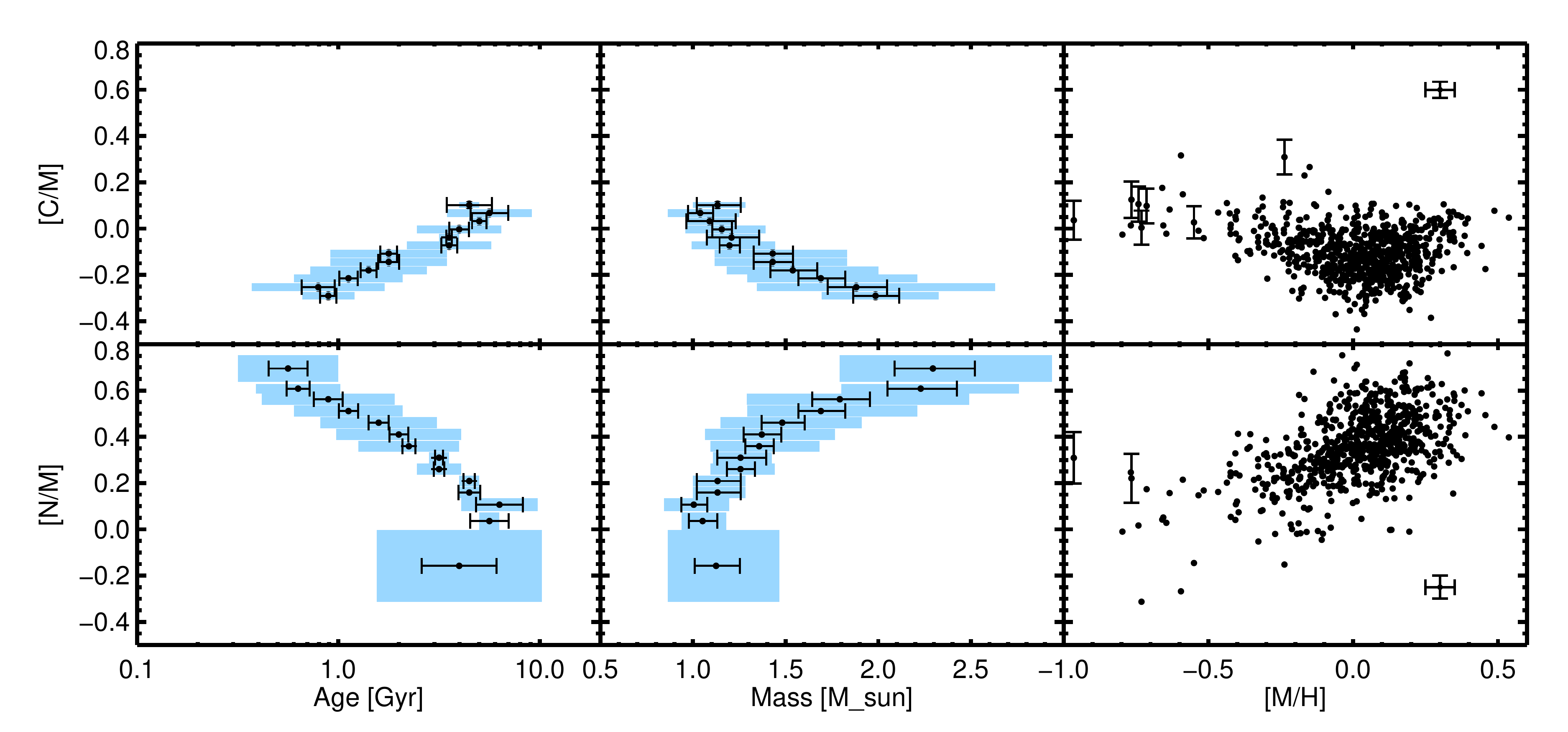} 
\caption{The left panels are the same as the left panel of Fig.~\ref{fig:age_al} but 
for [C/M] and [N/M]. The middle panels show the result of the hierarchical
modeling as a function of inferred
mass instead of age. The right panels show the [X/M]
vs [M/H] distributions for C and N. The typical uncertainty is shown in the
top right corner for C and bottom right corner for N. Stars for which the
[X/M] uncertainty is larger than twice the typical uncertainty have their
individual uncertainties shown.} 
\label{fig:age_c_n} 
\end{figure*}

\begin{figure}
\centering 
\includegraphics[clip, trim=5cm 2.5cm 1cm 2.8cm, width=0.47\textwidth]{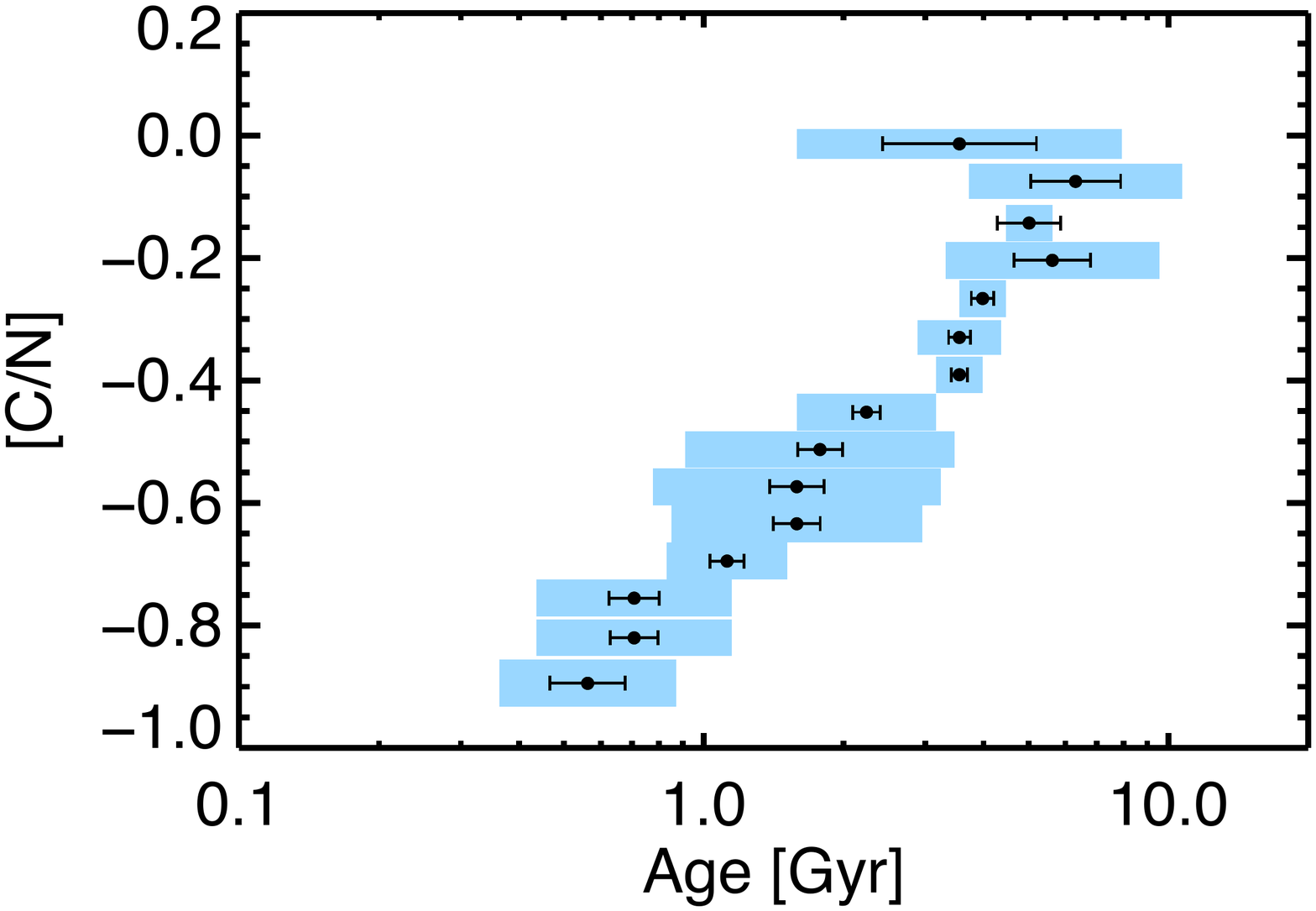}
\includegraphics[clip, trim=5cm 2.5cm 1cm 2.8cm, width=0.47\textwidth]{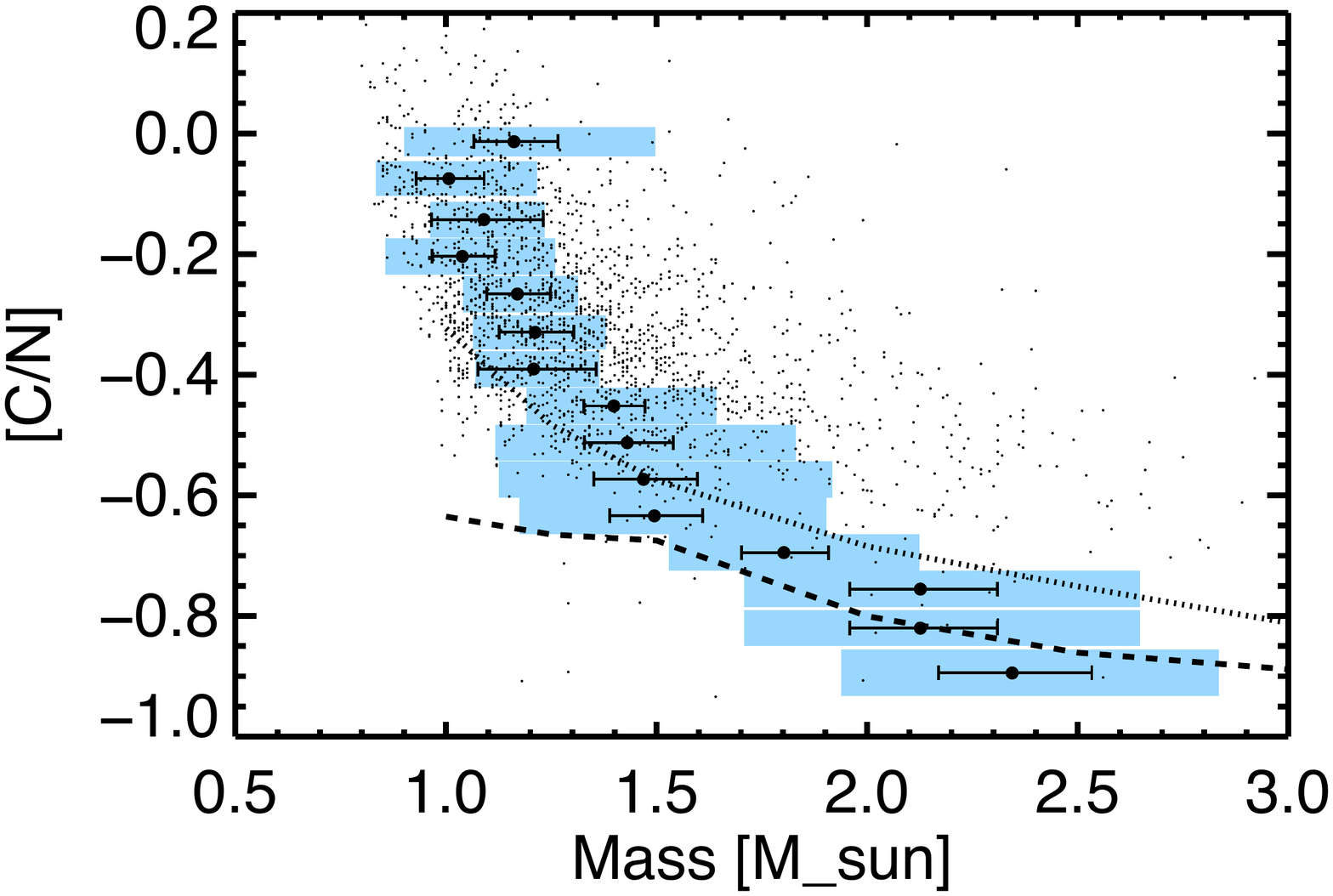} 
\caption{The top panel is the
same as the left panels of Fig.~\ref{fig:age_al} but for [C/N]. The bottom panel
shows the results of the hierarchical modelling as a function of 
inferred mass instead of age. Theoretical models from \citet{Lagarde2012} are shown with (dashed line)
and without (dotted line) rotation and thermohaline mixing. The small dots show the sample of APOKASC stars from \citet{Masseron2017a}} 
\label{fig:age_cn}
\end{figure}

\begin{figure}
\centering 
\includegraphics[clip, trim=5cm 2.5cm 1cm 2.8cm, width=0.47\textwidth]{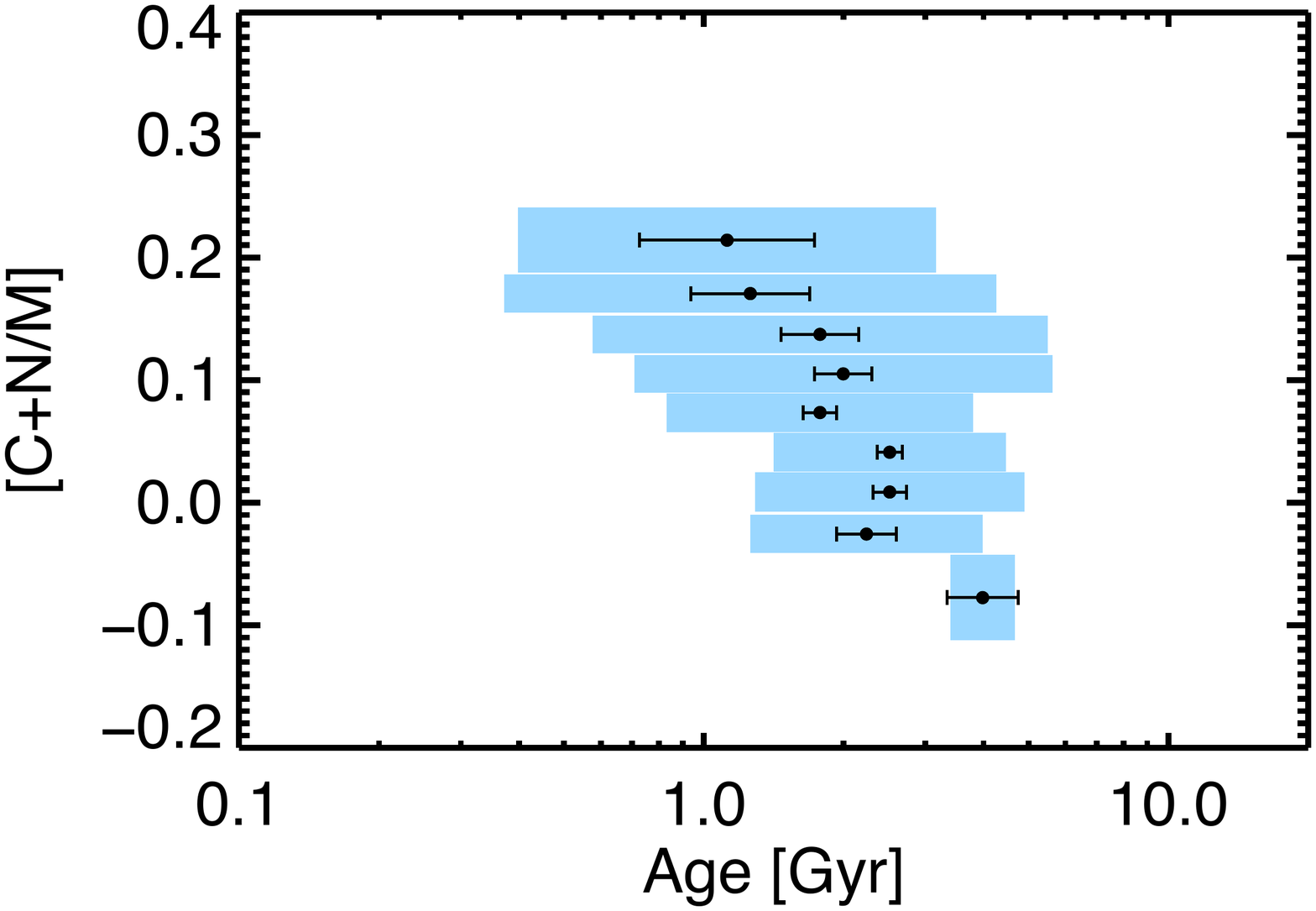}
\caption{The same as for the left panel of Fig.~\ref{fig:age_al} but for [C+N/M].} 
\label{fig:age_cnm}
\end{figure}

It is interesting to note that there is some discontinuity in the-
age--metallicity relation between [M/H] of $-0.4$ and $-0.2$ in
Fig.~\ref{fig:age_fe_o}. Four metallicity bins have ages younger than the
general trend suggests, and also lower age dispersions. At [M/H] of $-0.4$ the
low-\al\, sequence starts, and by [M/H] of $-0.2$, the low-\al\, sequence
strongly dominates the metallicity bin, see Fig.~\ref{fig:age_al}. These four
bins also have smaller age dispersions than the other metallicity bins,
suggesting that at [M/H] between $-0.4$ and $-0.2$, the stars in the high- and low-\al\,
sequences have comparable ages. Indeed, in Fig.~\ref{fig:age_alpha_fe}
the metallicity bin from $-0.3$ to $-0.2$ shows a very similar mean age at both
high and low [\al/M]. 
Again, we note that in our sample there are very few
\al-enhanced stars and more stars are needed to do a robust analysis on
single [\al/M], single [M/H] bins. The discontinuity at low abundances is not
strong in the [O/H] age trend, but the age dispersions in the bins around [O/H]
of $-0.3$ are also significantly smaller.

An important question when trying to interpret the observed [M/H]--age and [\al/M]--age relations is why the [M/H]--age relation is strongly affected by the possible radial migration, while the [\al/M]--age relation remains relatively steep and narrow. We include some discussion of this issue in \S \ref{sec:WAF}, but high-resolution models and larger observational samples with smaller age uncertainties are needed to fully address this issue.

The position of the Sun in the abundance--age relations of [\al/M], [M/H], and [O/H] (as well as most of the individual elements) is older than the age dispersion range of the solar abundance bin, assuming a solar age estimate of 4.66~Gyr from \citet{Dziembowski1999}. This may indicate that our age estimates are systematically too young. But this is only a systematic shift in the absolute age value and the relative ages are still reliable. Alternatively, this could mean that the Sun is a significant outlier in the solar neighborhood. The Sun would also lie on the older edge of the [\al/M]--age distribution of solar neighbourhood stars found by \citet{Haywood2013}. It is also older than the mean age of the solar bin in the [\al/M]--[M/H]--age figures of \citet{Martig2016}, \citet{Ho2017}, and \citet{Wu2017}, but these samples are much larger and extend beyond the solar neighborhood. 
The Sun as an outlier is not consistent with the [\al/M]--age relations found by \citet{Bensby2014} and \citet{Bergemann2014}, or with any age--metallicity relations from the literature.
The young B--star observations of \citep{Nieva2012} lend support to our observation that the youngest stars in the solar neighbourhood, are indeed those with solar metallicities. This suggests at least that the Sun was not born in the solar neighbourhood given its intermediate age and solar metallicity. Models of Galactic evolution have shown that the Sun has likely migrated 1--4 kpc from its birth radius \citep[e.g.][]{Wielen1996,Minchev2013}. However, other stars with similar ages and abundances to the Sun could have also migrated into the solar neighbourhood.

\subsection{Carbon and Nitrogen}

Carbon and nitrogen abundances in the atmospheres of red giant stars are
affected by internal mixing of material during the first dredge-up phase.
During this phase, the convective envelope extends into the H-burning region in
the core and brings enriched material to the stellar atmosphere
\citep{Iben1965}. This material from the core has been enriched through the
CN cycle, which results in a build up of N and a depletion of C.
The depth of the convective zone during the first dredge-up phase is dependent on the mass of the star,
resulting in decreasing atmospheric C to N ratios in higher mass red giants
(see \citet{Salaris2015} for a detailed explanation).

The left panels of Fig.~\ref{fig:age_c_n} show the parameters of a SFH model
constrained as a function of [C/M] and [N/M]. Because the C and N surface
abundance ratio in giants are actually a function of mass and only manifest as
a correlation with age due to the tight mass-age relation for red giant stars,
we also show the results of the hierarchical modelling as a function of 
inferred mass in
the middle panels. For each abundance ratio bin we compute the average mass
and mass dispersion of red giants
given the hierarchically modelled SFH and a Chabrier
lognormal IMF. As in F16, we assume that the metallicity distribution function
is flat within the [M/H] uncertainty. The right panels of Fig.~\ref{fig:age_c_n}
show the [C/M] and [N/M] vs [M/H] distributions for the sample. Because the
creation of N in the CN cycle is caused by the depletion of C, we also show
the results of the hierarchical modelling for the [C/N] abundance ratio as a
function of both age and mass in Fig.~\ref{fig:age_cn}.

\begin{figure*}
\centering 
\includegraphics[clip, angle=90, trim=0.4cm 0.8cm 0.9cm 0.8cm, width=0.9\textwidth]{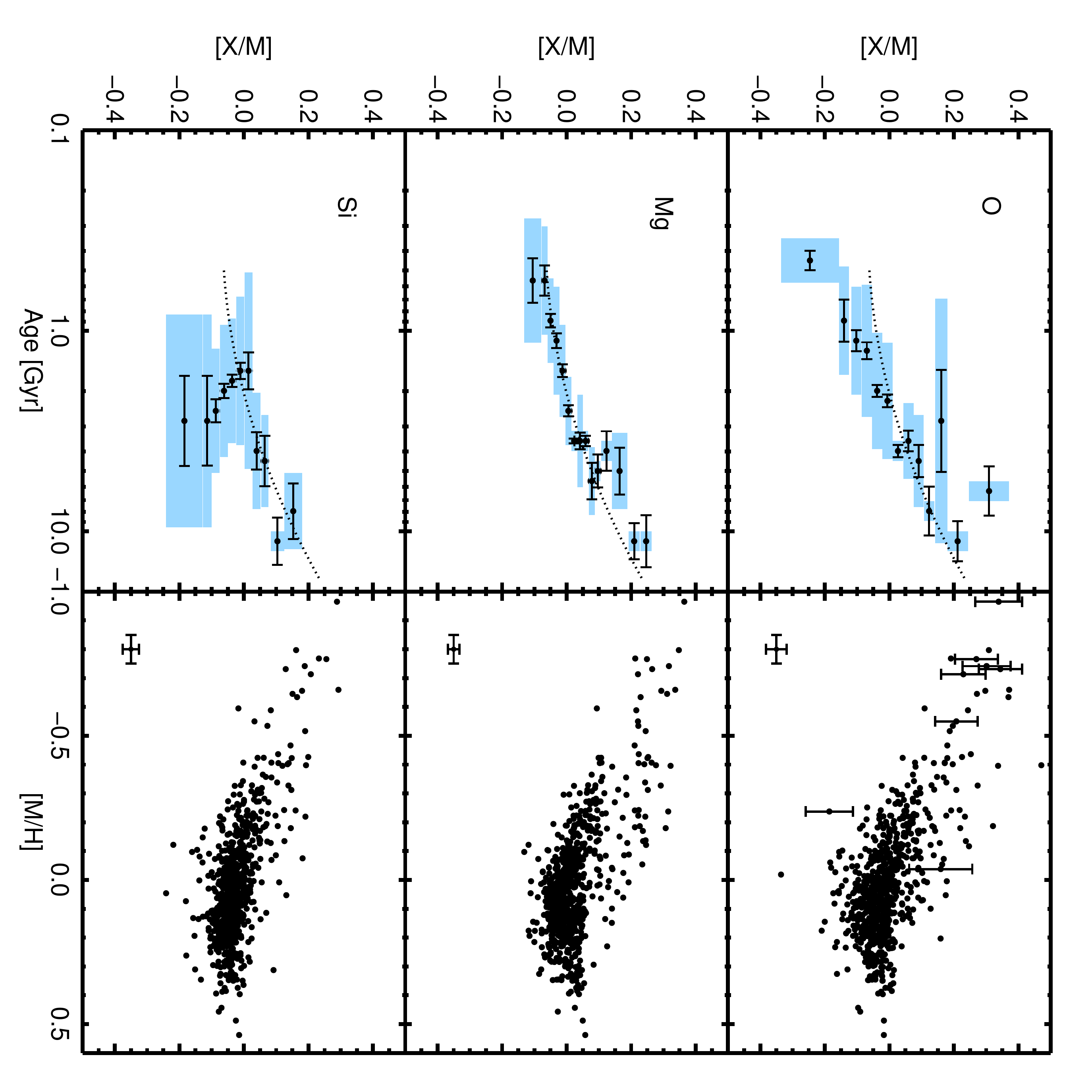} 
\caption{Same as Fig.~\ref{fig:age_c_n} for models of SFH constrained as a function of [X/M] for
O, Mg, and Si. As in Fig.~\ref{fig:age_al}, the dotted line represents the
quadratic fit to the \al-age relation. The typical uncertainty is shown in the
bottom left corner. Stars for which the
[X/M] uncertainty is larger than twice the typical uncertainty have their
individual uncertainties shown.
} 
\label{fig:age_OMgSi} 
\end{figure*}

We find a steep increase in age and decrease in mass with increasing [C/M], as
well as a contrasting steep decrease in age and increase in mass with
increasing [N/M]. The relation is also clearly seen in the [C/N] abundance
ratio. This is consistent with the CN cycle and dredge-up phase having a
stronger effect on higher mass, younger red giant stars. This effect has been
explored both with models and empirically as a means of separating stellar
populations by \citet{Masseron2015} and as a more precise proxy for individual
ages of red giants in the full APOGEE sample by \citet {Martig2016}. The
[C/N]-mass relation derived here compares very well with the relation presented
by \citet{Masseron2017a} using APOKASC (APOGEE + {\it Kepler} Asteroseismology
Science Consortium) targets, which have asteroseismic mass estimates. Fig.~\ref{fig:age_cn} shows the sample from \citet[][small black dots]{Masseron2017a} for comparison. Our
sample has slightly lower [C/N] values, but we use calibrated DR14 values
instead of externally corrected DR12 values as is done by
\citet{Masseron2017a}. We show results from the models of \citet{Lagarde2012}
in Fig.~\ref{fig:age_cn} with (dashed line) and without (dotted line) rotation
and thermohaline mixing. The comparison of our results to the models of
\citet{Lagarde2012} is consistent with that of \citet{Masseron2017a}, who do a
similar comparison using APOKASC clump stars in the open cluster NGC 6819, see
their fig.~5. Our results support the validity of the methods of
\citet{Martig2016} to use [C/M] and [N/M] abundance ratios as an age proxy for
giants in the solar neighbourhood.

The massive RC stars in our sample have preferentially high [N/M], low [C/M], and low [C/N]. Therefore the \logg\, overcorrection for massive RC stars discussed in \S \ref{sec:method} could be pushed the lowest [C/N] bins towards older ages and lower masses. If corrected, the mean mass of the lowest [C/N] bins would be more massive and the trend would flatten, bringing it into better agreement with both the models of \citet{Lagarde2012} and the APOKASC observations of \citet{Masseron2017a}.

Recent studies of solar twins in the solar neighborhood have found a trend in [C/Fe] with age such that solar twins with low [C/Fe] are younger than stars with high [C/Fe], although with a large amount of scatter \citep[][]{Nissen2015, Spina2016, Nissen2017, Bedell2018}. As these solar twin stars have not undergone internal mixing like the giant branch stars, this [C/Fe]-age relation is likely to reflect Galactic evolution rather than stellar evolution. If corrected for such an underlying trend, the observed [C/N]-mass relation would come into slightly better agreement with the models. Any correction based on the solar twins would be small, up to 0.1~dex in [C/Fe]. In order to further check for possible underlying trends of C and N with age, we examine the [C+N/M]-age relation, Fig.~\ref{fig:age_cnm}. The CN cycle creates an excess of N while depleting C, but should preserve the total C+N abundance. We find that the trend with age is fairly flat with a large age dispersion for most [C+N/M] abundances. There is a weak trend such that stars with high [C+N/M] are younger than stars with low [C+N/M]. This is opposite to the [C/Fe]-age trend observed in solar twins. Unfortunately, N is almost impossible to measure in optical spectra, so no [N/Fe]-age trend has been examined for solar twins. We find that potential corrections to the CN-mass relation of giants in the solar neighborhood due to underlying Galactic trends are small, but should be better characterised by looking at [N/M]-age trends in APOGEE dwarf and subgiant stars.

Although our results confirm the close relation between [C/N] and mass/age for
giants, we caution that this trend does have a mean log(age) dispersion of 0.2~dex. 
This sets a lower limit of approximately 0.2~dex, depending on the CN abundance uncertainties, 
on the precision with which the age of a giant can be determined 
from its [C/N] ratio. This limit is consistent with the age uncertainty reported 
by \citet{Martig2016} and \citet{Ness2016b}.
We caution that this is only a confirmation of the relation in the solar
neighbourhood and may not be applicable in other regions of the Galaxy. As
discussed by \citet{Martig2016}, differences in the birth abundances of C and N
at different Galactocentric radii could introduce false trends in the age.
Recent improvements in the modelling of stellar interiors has revealed that
including the thermohaline instability can have a significant effect on surface
[C/N] abundances of evolved stars \citep[see][]{Lagarde2017}, confirming the
need for great care when using surface C and N abundances to interpret age
trends of giant stars.

We emphasize again that the trends in Fig.~\ref{fig:age_c_n} and~\ref{fig:age_cn}
do not reflect Galactic chemical evolution like those in 
Fig.~\ref{fig:age_al}-\ref{fig:age_alpha_fe}. Rather, for a star to 
enter our sample it must be a red giant, and its age is therefore
tightly correlated with its mass. The mass in turn determines the
level of carbon depletion and nitrogen enhancement in the surface 
abundances of giant stars as a consequence of nuclear processing
and dredge up.


\subsection{The \al-elements} 
\label{sec:alpha_disc}

\begin{figure*}
\centering 
\includegraphics[clip, angle=90, trim=0.4cm 0.8cm 0.9cm 0.8cm, width=0.9\textwidth]{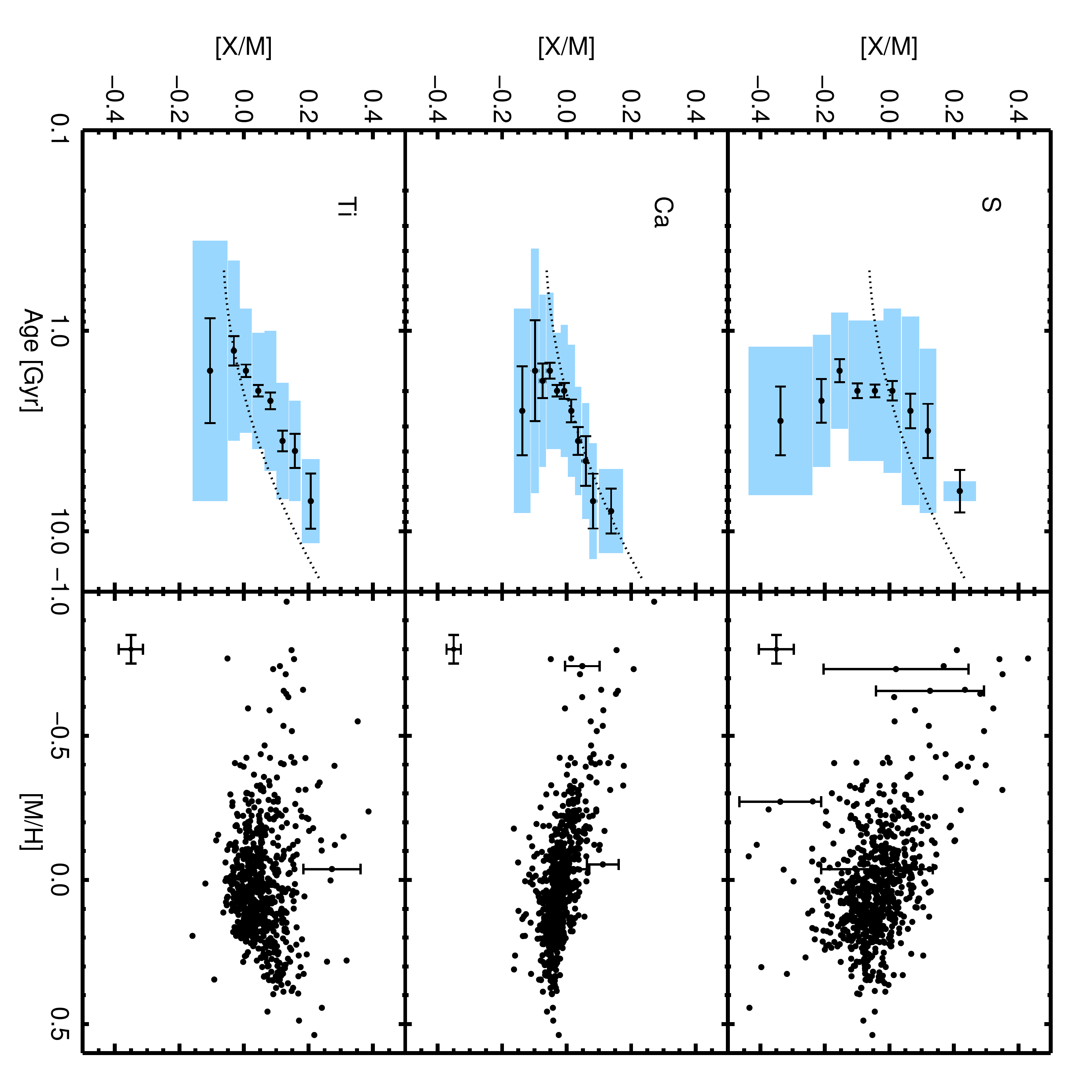} 
\caption{Same as
Fig.~\ref{fig:age_c_n} for models of SFH constrained as a function of [X/M] for
S, Ca, and Ti. As in Fig.~\ref{fig:age_al}, the dotted line represents the
quadratic fit to the \al-age relation.} 
\label{fig:age_SCaTi} 
\end{figure*}

The \al-elements measured by APOGEE are O, Mg, Si, S, Ca, and Ti.
Fig.~\ref{fig:age_OMgSi} and~\ref{fig:age_SCaTi} show the parameters of a model SFH
constrained as a function of these \al-elements and the [X/M] vs. [M/H]
distributions, with the typical uncertainty indicated in the top right corner
of each panel. We find generally that the individual \al-elements increase in
mean age with increasing [X/M] abundance ratio, in agreement with the overall
[\al/M] trend from Fig.~\ref{fig:age_al} and F16. The agreement with the dotted
line fit to the \al-age relation is extremely close for O, Mg, and the higher
abundance bins of Si and Ca.

{\it Oxygen.} We find that the mean age steeply increases with [O/M], and that [O/M]
strongly increases with decreasing [M/H]. These trends are in agreement with
many previous observations of dwarfs and subgiants in the solar neighbourhood
\citep[e.g.][]{Ramirez2013, Nissen2015, Spina2016, Adibekyan2016}. The bin
with [O/M] $\approx 0.15$ is a bit of an outlier with a younger mean age and
very large age dispersion compared to the overall trend. The stars in this bin
also cover a large range in metallicity, though this is true for
many other bins as well. As H. J\"onsson et al. (2018, in preparation) discuss, the ASPCAP [O/M] abundances are lower than literature results in the metal-poor stars. The difference in [O/M] between ASPCAP and literature appears to have a [M/H] dependence, which could be caused the strong sensitivity of the OH lines to \teff. The lowest [O/M] bins could be affected by the \logg overcorrection for massive RC stars. If corrected, these bin would be slightly younger, bringing the trend into better agreement with the [\al/M] line.

{\it Magnesium} The increase in mean age with [Mg/M] is strikingly well
defined, with a smaller age dispersion at each [Mg/M] bin than is seen in
[\al/M] and the other elements. This could be in part because of the high precision of the
Mg abundance measurements, see the typical uncertainty shown in Fig.~\ref{fig:age_OMgSi}. 
A strong [Mg/M]-age relation has been observed
in other samples of local dwarfs \citep[see e.g.][]{Nissen2015, Spina2016, Bedell2018}.
The tightness of the relation shown here is interesting because, as seen in
Fig.~\ref{fig:age_OMgSi}, stars with a single [Mg/M] abundance have up to a
0.8~dex spread in [M/H], but the age dispersion is small. This suggests a much
closer relation between age and [Mg/M] than [M/H]. This is consistent with the
SFH models constrained as a function of [M/H] in Fig.~\ref{fig:age_fe_o}. The
stars in each [M/H] bin have a range of [Mg/M] values as well as a large age
dispersion. From comparisons to independent analyses and literature results, H. J\"onsson et al. (2018, in preparation) conclude that Mg is the most accurately determined \al-element abundance in APOGEE. The lowest [Mg/M] bin could be affected by the \logg overcorrection for massive RC stars. If corrected, this bin would be slightly younger, bringing the trend into better agreement with the [\al/M] line.

{\it Silicon.} The mean age increases with increasing [Si/M] above solar, but
below solar values the mean age decreases with increasing [Si/M] and has a
larger age dispersion. The lower [Si/M] bins are in strong disagreement with
the [\al/M] trend and literature results. Previous studies of solar-like stars
in the solar neighbourhood have found a smooth increase in age with increasing
[Si/Fe], and specifically young ages for low [Si/Fe] stars \citep[see
e.g.][]{Nissen2015, Spina2016, Bedell2018}. \citet{Snaith2014} find a steep and tight relation 
between [Si/Fe] and age for Si enhanced dwarfs in the solar neighbourhood that 
flattens but continues at young ages and low [Si/M]. The shape of
the [Si/M] with [M/H] of our sample is consistent with the solar neighbourhood
sample of \citet{Bensby2014}, although the values are a bit lower in APOGEE. As noted by J\"onsson et al. (2018, in preparation), the APOGEE DR14 [Si/M] vs [M/H] is now consistent with the analysis of \citet{Hawkins2016}. Therefore, we do not find a convincing reason to distrust the APOGEE DR14 Si abundances in this sample despite the odd age trend.

\begin{figure*}
\centering 
\includegraphics[clip, angle=90, trim=0.4cm 0.8cm 0.9cm 0.8cm, width=0.9\textwidth]{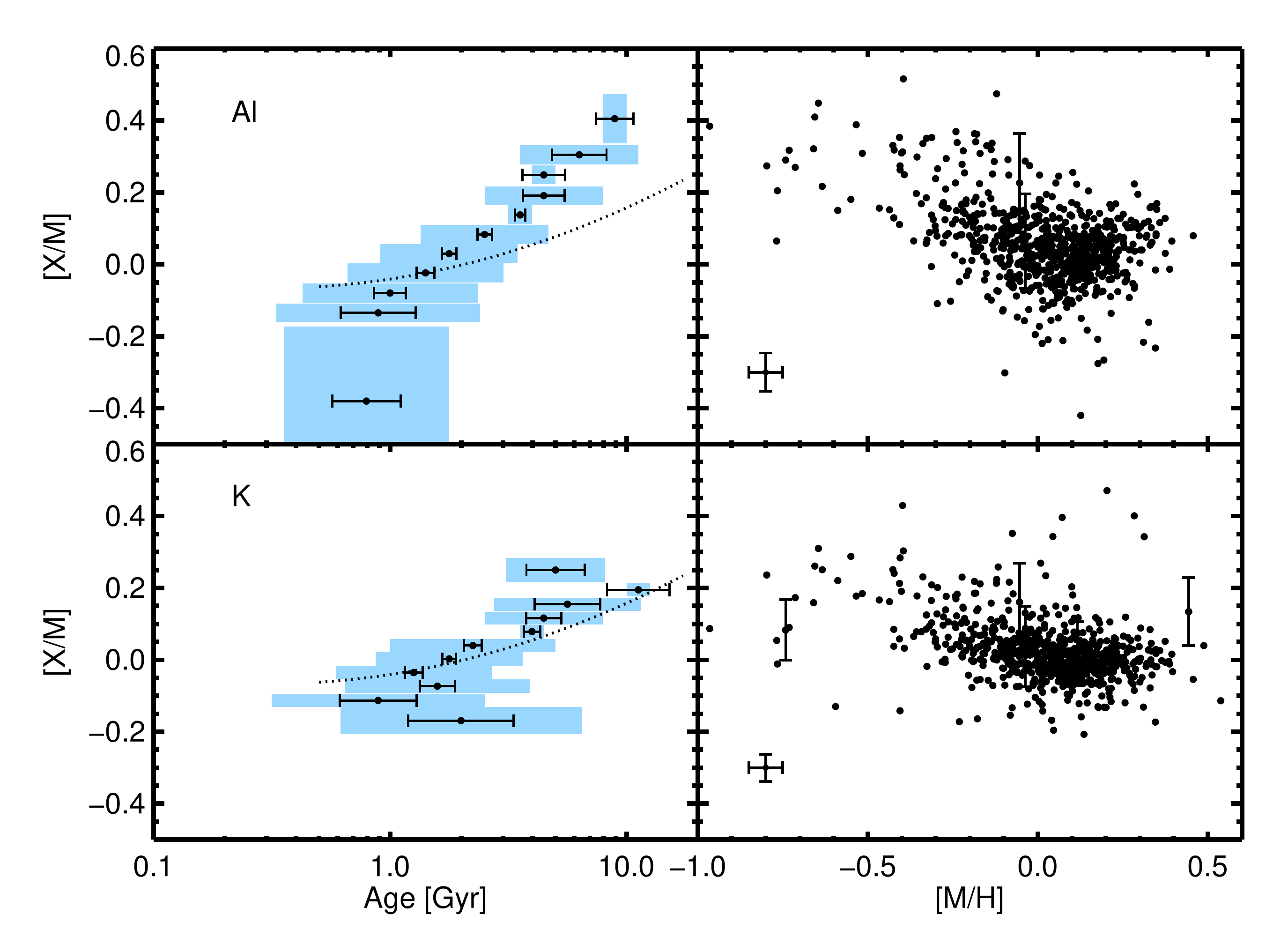} 
\caption{Same as Fig.~\ref{fig:age_OMgSi} but for models of SFH constrained as a function of [X/M] for
Al and K. As in Fig.~\ref{fig:age_al}, the dotted line represents the quadratic
fit to the \al-age relation.} 
\label{fig:age_AlK} 
\end{figure*}

{\it Sulphur.} We find that S behaves similarly to Si, but is consistent with a flat relation. 
The abundance bins are larger due to larger
abundance uncertainties and the age dispersion is large at all [S/M].
\citet{Spina2016} and \citet{Adibekyan2016} both find a flat relation at lower
[S/Fe] abundances and younger ages, but an increase in age for the most [S/Fe]
rich stars. \citet{Nissen2015} finds a steady increase in the age with
increasing [S/Fe], although they have no stars with the age of the most [S/Fe]
rich stars of \citet{Spina2016} and \citet{Adibekyan2016}. Although there is
an overall increase in the [S/M] abundance with decreasing metallicity, there
is a large spread in the [S/M] vs. [M/H] distribution and the typical APOGEE
uncertainty in [S/M] is larger than many of the other elements. The large
[S/M] measurement uncertainties could be blurring the age trend, see Fig.~\ref{fig:ab_err}.

{\it Calcium.} Ca shows an increase in mean age with increasing [Ca/M], which
steepens at higher [Ca/M] abundances. The lower [Ca/M] bins have stars with a
larger spread in [M/H] and also larger age dispersions. The lowest [Ca/M] bin
has an older mean age and a very large age dispersion, in the same manner as
the low [Si/M] and [S/M] bins. However, as this is the lowest abundance bin,
we are cautious to draw strong conclusions from this bin alone. \citet{DaSilva2012} and
\citet{Spina2016} find that [Ca/Fe] decreases with age for young stars, but
increases with age after about 5~Gyr. Our result of a steeper increase in mean
age at the most Ca-enhanced bins, and perhaps a turn towards older ages in
the lowest [Ca/M] bin, is consistent with these results. \citet{Nissen2015}
and \citet{Adibekyan2016} find the [Ca/M]-age relation to fairly flat, although they do not sample a large range of [Ca/M] values.

{\it Titanium.} We find an increase in age with increasing [Ti/M], similar to
the [\al/M] trend, with slightly younger mean ages. Although there is a large
age dispersion in all [Ti/M] bins, particularly the lowest bins, the trend in
mean age is surprisingly smooth given the [Ti/M] vs. [M/H] distribution. As
has been pointed out by \citet{Holtzman2015} and \citet{Hawkins2016}, the
APOGEE Ti abundances are known to have a different [Ti/M] vs [M/H] distribution
than the other \al-elements, in disagreement with other detailed abundance
studies of the Milky Way disc \citep[see e.g.][]{Bensby2014}.
\citet{Hawkins2016} find that this disagreement is driven by the use of two Ti
lines in ASPCAP that are either saturated or suffer from non-LTE effects.
J\"onsson et al. (2018, in preparation) determine that the Ti disagreement is likely due to
the use of uncalibrated instead of calibrated atmospheric parameters,
particularly \teff, by ASPCAP when determining
elemental abundances. Despite this disagreement in [Ti/M] vs [M/H], our age
trend with [Ti/M] is consistent with previous studies \citep{DaSilva2012,
Nissen2015, Spina2016, Adibekyan2016}. We do not recommend drawing strong
conclusions from the Ti results given the known issues with ASPCAP Ti
abundances \citep[see][]{Holtzman2015}.

While the \al-elements generally increase in age with increasing [X/M]
abundance ratio, Si, S, and perhaps Ca show a turn towards older ages in the
lowest [X/M] bins. The stars with the lowest [O/M], [Mg/M], and [Ti/M]
abundance ratios are not the most metal-rich stars, showing the typical
`banana' shape of the low [\al/M] sequence. However, the stars with the lowest
[Si/M], [S/M], and [Ca/M] abundance ratios do not have a `banana' shape, but
show a decreasing or flat abundance ratio with metallicity even in the most
metal-rich stars. These metal-rich stars were seen to have older ages than the
solar-metallicity stars in Fig.~\ref{fig:age_fe_o}. Si, S, and Ca are also
theoretically expected to have larger SNIa contributions than Mg and O
\citep[e.g.][]{Thielemann2003}, which could explain their different behavior.

\subsection{Light odd-Z elements}

Fig.~\ref{fig:age_AlK} gives the age trends and [X/M] vs.
[M/H] distributions for the light odd-Z elements Al, and K. Both [Al/M]
and [K/M] strongly increase in age with increasing abundance ratio. We briefly discuss the trends found for Na and P, although they are not shown because the APOGEE abundance uncertainties are too large for a meaningful examination of the age-abundance trends. A full discussion of the effects of abundance uncertainties on the age trends is given in \S \ref{sec:uncertainties}. 

{\it Aluminum.} The mean age smoothly increases with increasing [Al/M]. The
[Al/M]-age trend is similar to, but steeper than the [\al/M]-age relation line.
The [Al/M] vs [M/H] distribution is similar to the \al-elements but with more scatter.
This is consistent with the observations of \citet{Nissen2015},
\citet{Spina2016}, and \citet{Bedell2018} who find that [Al/Fe] steeply increases with age.
\citet{Smiljanic2016} find a shallower [Al/Fe]-age relation using GES data.
\citet{Bensby2014} also find an \al-like behavior of [Al/Fe] with large scatter at
the metal-poor end. \citet{Hawkins2016} found that some of the Al~{\footnotesize I} lines are 
saturated or likely suffer from NLTE effects. \citet{Nordlander2017} find that the 16750~\AA\, Al~{\footnotesize I} line is affected by departures from LTE and also from hyperfine splitting.

{\it Potassium.} We find that the mean age increases steadily with increasing
[K/M], very similar to the \al-age relation line. The [K/M] vs [M/H]
distribution is also \al-like. The bin with the highest [K/M] abundance ratio
could be strongly influenced by the metal-rich, K-enhanced stars. Like Ca,
the lowest [K/M] bin has an older mean age and the stars in this bin have a
1.0~dex range in metallicity. These highest and lowest [K/M] bins contain
several stars that do not lie within the main [K/M] vs [M/H] distribution. 
The K lines used by ASPCAP are both slightly blended.
\citet{Spina2016} do present K abundances, but they do not find any
correlations between age and [K/Fe], having very large scatter in [K/Fe] at all
ages.

\begin{figure*}
\centering 
\includegraphics[clip, angle=90, trim=0.4cm 0.8cm 0.9cm 0.8cm, width=0.9\textwidth]{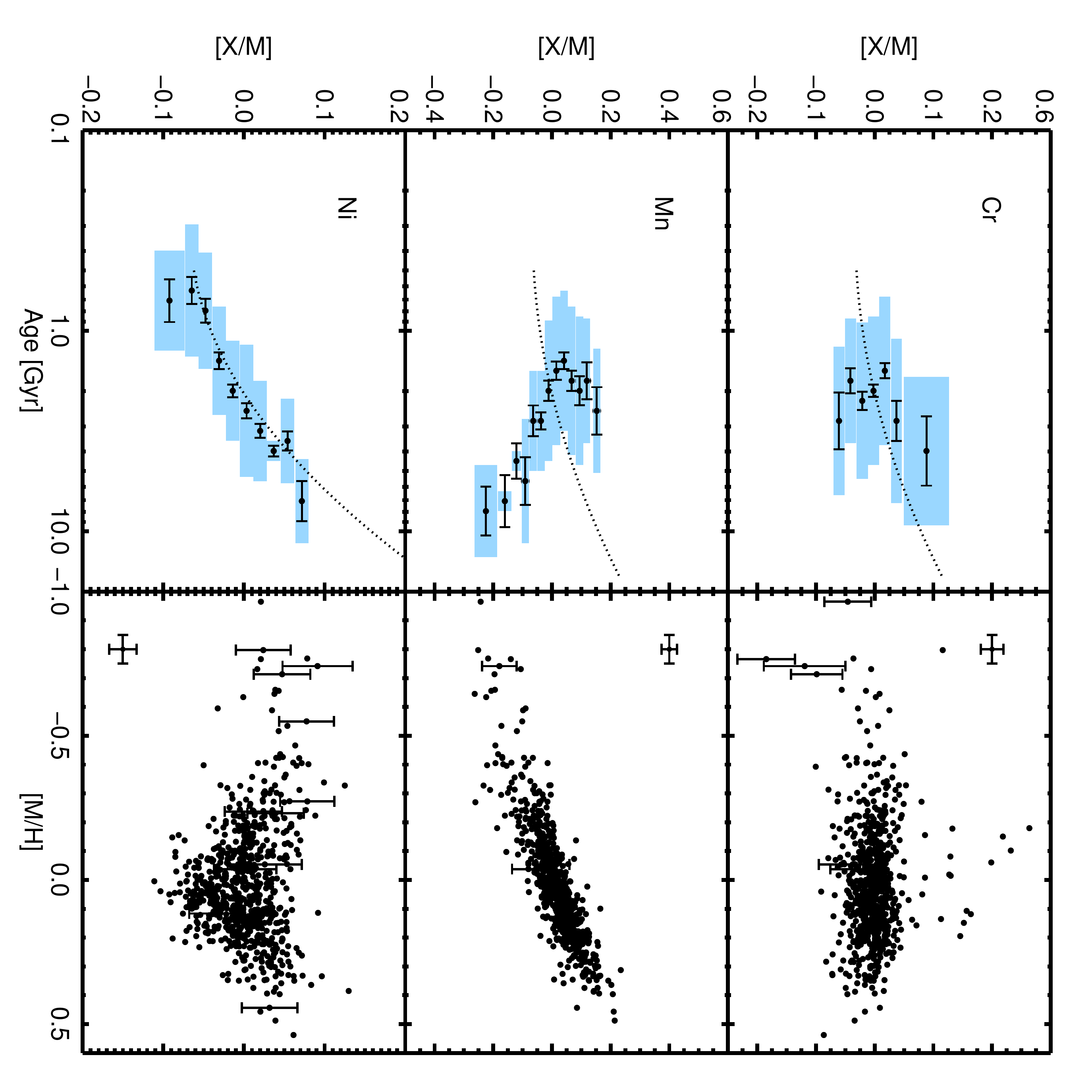} 
\caption{Same as Fig.~\ref{fig:age_OMgSi} but for models of SFH constrained as a function of [X/M] for
Cr, Mn, and Ni. As in Fig.~\ref{fig:age_al}, the dotted line represents the fit
to the \al-age relation.} 
\label{fig:age_CrMnNi} 
\end{figure*}

{\it Sodium.} We find that the mean age slightly decreases with increasing
[Na/M], with a large age dispersion at most abundances. The [Na/M] vs [M/H]
distribution is very scattered, but is generally increasing in [Na/M] with
increasing [M/H]. The uncertainties in Na are large 
because the two Na lines available in the APOGEE spectra are very weak.

{\it Phosphorus.} The mean age is slightly decreasing with increasing [P/M],
with large age dispersions in all bins, similar to the behavior of [Na/M]. The
[P/M] vs. [M/H] distribution appears to be generally decreasing in [P/M] with
increasing metallicity, although the scatter is very large. This is opposite
to the behavior of [Na/M].

\subsection{Fe-peak elements} 
\label{sec:fepeak_disc}

The age trends and [X/M] vs [M/H] distributions for the Fe-peak elements are
shown in Fig.~\ref{fig:age_CrMnNi}. The
measurement of [M/H] in ASPCAP is heavily weighted by the contributions from
elements in this group. The behavior of [Ni/M] is remarkably \al-like, while the behavior 
of [Mn/M] closest matches the [M/H] trend. V and Co are not shown due to their 
larger abundance uncertainties.

{\it Vanadium.} We find that the mean age is flat with [V/M], with a large age
dispersion at all abundance bins. The [V/M] vs [M/H] distribution
has large scatter with no clear trend and large uncertainties. 

{\it Chromium.} The mean age is flat with [Cr/M] over the narrow range of
[Cr/M] abundances, with a large age dispersion at all [Cr/M] bins. In this
sample, [Cr/M] is flat with metallicity, in agreement with \citet{Bensby2014}.
There are a few high [Cr/M] stars around solar metallicity and one with very
low metallicity that do not lie within the main [Cr/M] vs [M/H] distribution.
These highest [Cr/M] stars have a slightly older mean age. Recent studies of
solar-like stars find [Cr/M] to be either flat or weakly decreasing with age
\citep{DaSilva2012, Nissen2015, Spina2016, Adibekyan2016, Bedell2018}, which is consistent
with our results.

{\it Manganese.} The abundance-age relation of [Mn/M] is very similar to the
age-metallicity relation from Fig.~\ref{fig:age_fe_o}. We find that the mean age
decreases with increasing [Mn/M] until about solar [Mn/M], then increases with
increasing [Mn/M] abundance above solar [Mn/M]. The [Mn/M] vs [M/H]
distribution is very tight, and smoothly increases with increasing [M/H] across
all metallicities. Again, this is consistent with the most metal-rich stars
being older than the solar abundance stars. This age trend is consistent with
\citet{DaSilva2012} who find that the highest [Mn/M] stars have intermediate
ages. \citet{Spina2016} find large scatter in the [Mn/M]-age relation.

This [Mn/M] vs. [M/H] distribution is very similar to the distribution found
by \citet{Battistini2015} without applying NLTE corrections.
\citet{Battistini2015} find that the distribution looks quite different after
applying NLTE corrections, becoming a flat distribution. This brings it into agreement with
the Ni and Cr distributions. Although that analysis was done with optical
lines, the strong agreement between APOGEE and the LTE distribution of
\citet{Battistini2015} is an indication that there could be NLTE corrections
needed for the $H$-band Mn lines used by APOGEE. However, it may also indicate that 
the NLTE correction for Mn is overestimated for the optical lines.

{\it Cobalt.} We find a large amount of scatter in the mean age of the [Co/M]
bins, with no significant trend with abundance. There are also large age
dispersions at all [Co/M] bins. The [Co/M] vs [M/H] distribution has a large
scatter as well, with a large mean uncertainty. 

{\it Nickel.} Although the range of [Ni/Fe] abundances is small (note the
change in $y$-axis range for Ni in Fig.~\ref{fig:age_CrMnNi}), the ASPCAP
uncertainties are also very small. The mean age increases steeply and smoothly
with increasing [Ni/M] over the narrow abundance range. The agreement with the
\al-age line is remarkably good. The [Ni/M] vs [M/H] distribution is also very
tight and looks \al-like but flatter. The age trend is consistent with
previous studies that find an overall increase in age with increasing [Ni/Fe]
abundance \citep[e.g.][]{Nissen2015, Nissen2017, Spina2016, Bedell2018}. However, the trend presented
here is steeper than in previous work. Although the Ni trends agree with observed trends on solar twins, the strong \al-like behavior of Ni is not predicted by theoretical nucleosynthesis calculations \citep[e.g.][]{Kobayashi2006}.

\subsection{Effects of abundance uncertainties}
\label{sec:uncertainties}

\begin{figure*}
\centering 
\includegraphics[clip, angle=90, trim=0.5cm 1cm 1cm 1cm, page=1, width=1.0\textwidth]{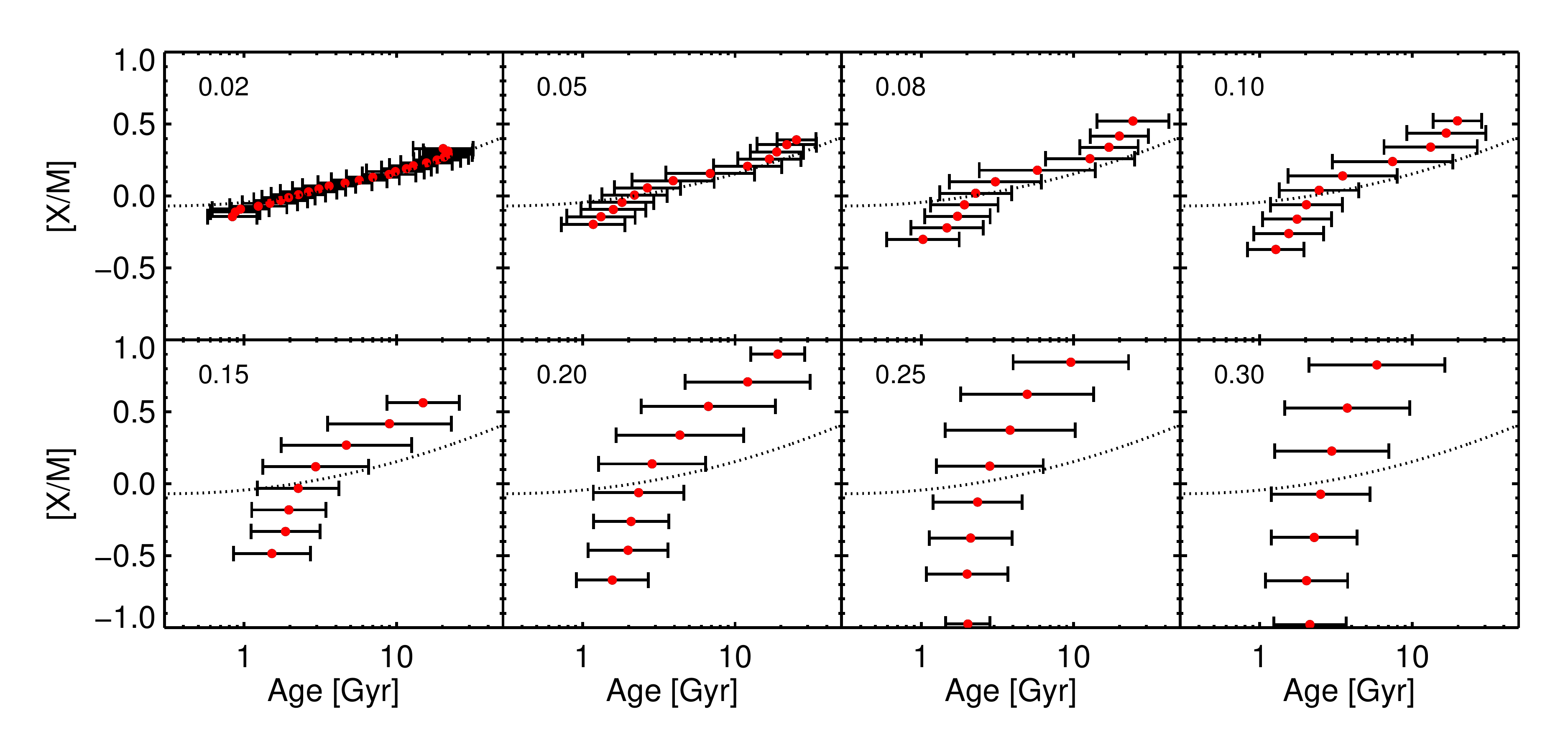} 
\caption{The abundance--age relations for mock data with age uncertainties of 0.16 dex and increasing abundance uncertainties from 0.02 -- 0.30 dex. The red points are the mean age of the abundance bin and the error bars show the standard deviation of the mean age. The observed \al--age fit, Equation \ref{eq:alpha_fit} is shown as the dotted line. The [X/M] uncertainty of each panel is shown in the upper left corner of the panel.} 
\label{fig:ab_err} 
\end{figure*}

We do note that uncertainties in the ages and elemental abundance measurements can have the effect of blurring the presence of an abundance-age relation. As a test of this effect, we created from a mock sample of 10,000 stars randomly assigned [X/M] abundances to roughly match the [\al/M] distribution of our observed sample, i.e. more stars with low [X/M] than with high [X/M]. We performed the same test on a sample with a uniform distribution of [X/M] and find similar results. For each star we assigned an age determined using the quadratic fit to the observed [\al/M]--log(age) relation, Equation~\ref{eq:alpha_fit}. We then applied Gaussian noise to both the age and the [X/M] abundance with a $\sigma$ of 0.16~dex in log(age) and increasing from 0.02 to 0.30~dex in [X/M]. In Fig.~\ref{fig:ab_err} the resulting abundance-age trends binned in the same way as the observed data, i.e. bin widths equal to the abundance uncertainty and containing at least 15 stars. The red points indicate the mean age of the mock stars in the bin and the error bars indicate the standard deviation. Equation~\ref{eq:alpha_fit} is also shown as the dotted line to indicate the underlying [X/M]-age relation. The [X/M] uncertainty used in each panel is given in the upper left corner. 

From this test we find that an underlying abundance--age trend can be nearly completely erased if the abundance measurement errors are large enough. This could be strongly affecting our observed trends for Na, P, V, and Co, which have mean uncertainties of 0.09, 0.10, 0.07, and 0.09~dex, respectively. From Fig.~\ref{fig:ab_err} it appears that any underlying abundance--age relation in these elements must be shallower than the observed [\al/M]--age relation to result in their respective flat relations, given the observational uncertainties. Although the abundance uncertainties could explain the flat relations seen in Na, P, V, and Co, we do not find that larger abundance errors are able to produce the reversal in the abundance-age relation at either high or low [X/M] such as that observed in Si, S, Ca, and Mn (0.03, 0.05, 0.02, and 0.03~dex uncertainty, respectively). While the [S/M]--age relation looks similar to the trends produced by very large abundance uncertainties, the reported ASPCAP uncertainty is at least 0.10~dex smaller. This suggests that either the underlying [S/M]--age relation is significantly shallower than the observed [\al/M]--age relation or the ASPCAP uncertainties are significantly underestimated for this element. 

[O/M] and [Mg/M] are consistent with the observed [\al/M]--age relation within their mean uncertainties of 0.03 and 0.02~dex, respectively. However, given these uncertainties, the slight age offset of the high abundance ratio bins are significant and may indicate a parallel but offset abundance--age relation for the \al-rich sequence. The mean uncertainties of Al and K (0.05 and 0.04~dex, respectively) suggest that the underlying abundance--age relation may be stronger than the observed relation for these elements. The observed abundance--age relations of Cr, Mn, and Ni should also be consistent with the underlying trend given the uncertainties of 0.04, 0.03, and 0.02, respectively. However, it should be noted that we have only tested the effects of the abundance uncertainties assuming an \al-like age relation. While we did confirm that both a shallower age relation and larger age uncertainties also could not produce the [M/H]-like age relation, we did not have a robust [M/H]--age relation unaffected by possible migration with which to test the Fe-peak elements.

Through examining the abundance-age trends of the individual elements, we highlight that the differences between individual elements can be a tool used to disentangle the complicated processes of Galactic evolution. For example, we find surprisingly large variations in the abundance-age trends even among the \al--elements that cannot be explained by the age or abundance uncertainties alone. 

\section{Galactic Chemical Evolution Models} 
\label{sec:GCE}

A key goal of deriving age measurements and detailed chemical
abundance observations is to place constraints on models of Galactic and
chemical evolution. In this section we compare our observations to
the predictions of two relatively simple GCE frameworks, one 
analytic and one numerical. The successes and failures of these
models are both informative.

\subsection{One-Zone Models and Mixtures} 
\label{sec:WAF}

We first discuss our results in the context of simple one-zone chemical
evolution models and mixtures of such models that could represent the impact of
radial mixing on a disc whose evolutionary parameters vary with Galactocentric
radius. For this discussion we draw on analytic results from \citet*[][hereafter WAF17]{Weinberg2017}. Our goal is
not to derive a model that reproduces all aspects of the data but rather to
understand how physical parameters affect the abundance-age relations and
explore qualitative implications of the APOGEE measurements for disc history.

\begin{figure*}
\centerline{\includegraphics[width=5.5truein]{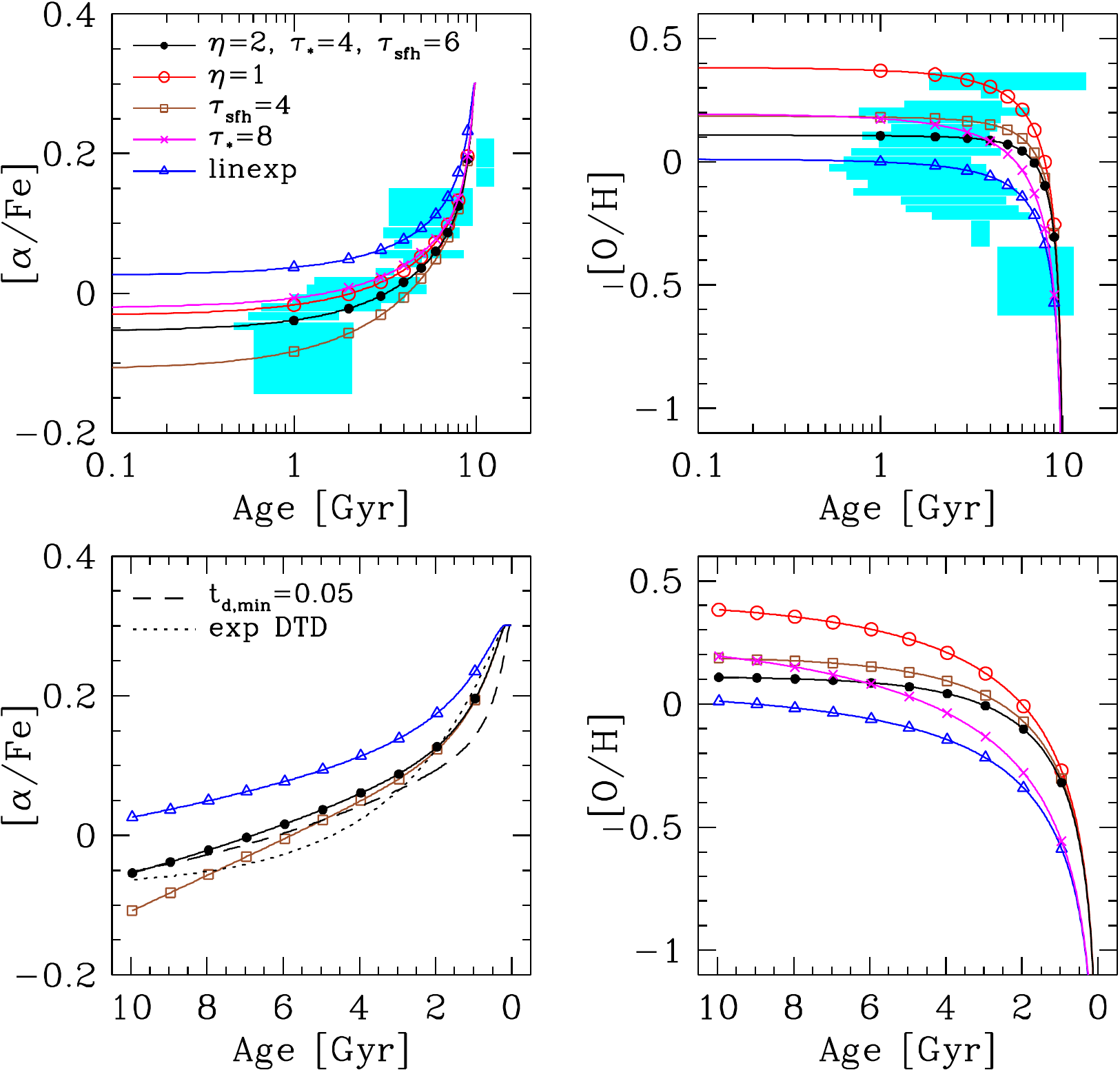}}
\caption{Evolutionary tracks of $\afe$ (left) and $\ohh$ (right) for five
one-zone models. In the upper panels cyan rectangles represent the
observational measurements from Fig.~\ref{fig:age_al} and~\ref{fig:age_fe_o}. The
black curves represents a model with $\eta=2$, $\tau_*=4\Gyr$, $\tausfh=6\Gyr$
and an exponential SFR. Filled circles mark abundances at $1\Gyr$ intervals.
Other curves show single-parameter variations away from this reference model: a
lower outflow mass-loading factor ($\eta=1$, red curve, open circles), a more
rapidly declining SFH ($\tausfh=4\Gyr$, sienna curve, squares), a lower star
formation efficiency ($\taustar=8\Gyr$, magenta curve, crosses), or a
linear--exponential SFH with the same parameter values (blue curve, triangles).
Solid curves in the lower panels show the same models with a linear instead
of logarithmic age axis.
In the lower left panel, two models have been omitted and replaced with two
variants of the central model, which have a minimum SNIa delay time of 0.05~Gyr
instead of 0.15~Gyr (dashed line) or a single-exponential DTD (dotted line). }
\label{fig:onezone} 
\end{figure*}

\begin{figure}
\centerline{\includegraphics[width=3.3truein]{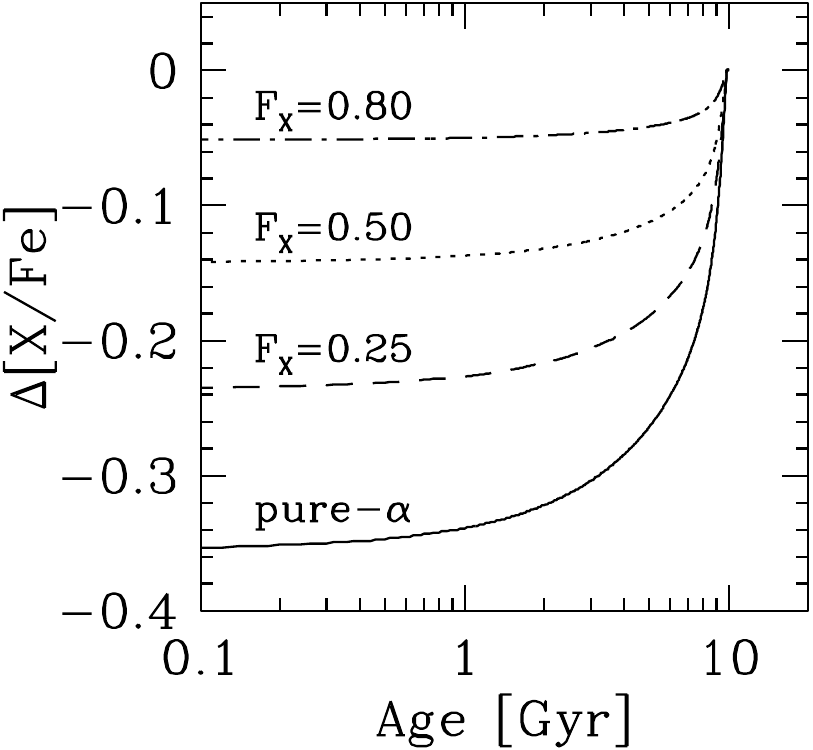}}
\caption{Predicted evolutionary tracks for elements with different
ratios of SNIa to CCSN contribution, but metallicity-independent yields,
in a one-zone model with the parameters of the black curve in 
Fig.~\ref{fig:onezone} ($\eta=2$, $\taustar=4$, $\tausfh=6$).
$\Delta\xfe$ is the change in $\xfe$ relative to the plateau value
implied by CCSN yields. $F_X$ is the SNIa contribution relative to
that of iron (see eq.~\ref{eqn:fx}), so a ``pure-$\alpha$'' element
corresponds to $F_X=0$.
}
\label{fig:fx} 
\end{figure}

The key parameters of the WAF17 analytic solutions are the IMF-averaged supernova
yields for core collapse supernovae (CCSN) and Type Ia supernovae (SNIa), the
star formation efficiency time-scale $\taustar \equiv \mgas/\mdotstar$, the
outflow mass-loading factor $\eta \equiv \mdotout/\mdotstar$, and a time-scale
for the star formation history $\tausfh$. The analytic solutions assume
metallicity-independent yields and a star formation history (SFH) that is
exponential, $\mdotstar(t) \propto e^{-t/\tausfh}$, linear--exponential,
$\mdotstar(t) \propto t e^{-t/\tausfh}$, constant, or a linear combination of
these forms. CCSN are assumed to return newly produced oxygen and iron to the
ISM instantaneously at a rate $\mocc \mdotstar$ and $\mfecc \mdotstar$,
respectively, while SNIa return iron at a rate $\mfeIa\mdotstar$ with
exponential delay time distributions (DTD). For the calculations here we use a
sum of two exponentials that approximates a $t^{-1.1}$ power-law DTD
\citep*{maoz2012} with a minimum delay time of 0.15~Gyr (see WAF17 for details).
Stars also instantaneously recycle a fraction $r$ of the gas and metals they
are born with; we adopt $r=0.4$, which is a good approximation for a
\cite{kroupa2001} IMF. In the absence of infall, the combination of star
formation and outflow would deplete the gas on an $e$-folding time-scale
\begin{equation} 
\taudep = {\taustar \over 1+\eta-r}~. 
\label{eqn:taudep}
\end{equation}

For an exponentially declining SFH, the evolution equation for the ISM oxygen
mass fraction $\Zo$ is 
\begin{equation} 
\Zo(t) = \Zoeq
\left(1-e^{-t/\taubar}\right)~, 
\label{eqn:ZoEvol} 
\end{equation} 
where the equilibrium abundance is 
\begin{equation} 
\Zoeq = {\mocc \over 1+\eta-r} \times
\left(1-\taudep/\tausfh\right)^{-1} 
\label{eqn:ZoEq} 
\end{equation} 
and the evolution time-scale is 
\begin{equation} 
\taubar = \left(\taudep^{-1}-\tausfh^{-1}\right)^{-1} =
 \taudep \times \left(1-\taudep/\tausfh\right)^{-1}~.
\label{eqn:taubar} 
\end{equation} 
For a linear--exponential SFH, the equilibrium
abundance and time-scale $\taubar$ are the same, but the functional form of the
solution yields a slower approach to equilibrium, 
\begin{equation} 
\Zo(t) = \Zoeq
 \left[1-{\taubar\over t}\left(1-e^{-t/\taubar}\right)\right]~.
\label{eqn:ZoEvolLinExp} 
\end{equation} 
With an appropriate substitution for $\mocc$, these solutions
describe the CCSN contribution to any element with a metallicity-independent
yield. We adopt oxygen as our representative $\alpha$-element and identify
$\ofe$ with $\afe$. Analytic solutions for the SNIa iron contribution can be
found in WAF17.

The gas infall rate is determined implicitly by the assumed star formation
efficiency and SFH (WAF17, eq. 9). For an exponential SFH one can easily show
that 
\begin{equation} 
\left(1-\taudep/\tausfh\right) = {\mdotinf \over
\mgas/\taudep}~, 
\label{eqn:taufactor} 
\end{equation} 
so the second factor in
equations~(\ref{eqn:ZoEq}) and~(\ref{eqn:taubar}) approaches unity if the
infall rate is sufficient to replenish the ISM at the rate it is being
depleted. For a gas-starved system with $\mdotinf \ll M_{\rm gas}/\taudep$,
the equilibrium abundance diverges and the time-scale for reaching it also
diverges; evolution in this case approaches the traditional ``closed box'' or
``leaky box'' analytic models. Conversely, the limit $\tausfh \rightarrow
\infty$ corresponds to constant SFR, with infall keeping the gas supply
constant and $(1-\taudep/\tausfh)=1$.

Fig.~\ref{fig:onezone} shows the evolutionary tracks of five one-zone models
in the $\afe$-log(age) plane and the $\ohh$-log(age) plane. For oxygen and
iron yields we adopt $\mocc=0.014$, $\mfecc=0.0015$, $\mfeIa=0.0015$, based
approximately on \cite{andrews2017} but adjusted slightly to better fit the
$\afe-\feh$ tracks measured by APOGEE. All models run for 10~Gyr, and points
along the model curves mark abundances at 1~Gyr intervals. One can see at a
glance that most of these models approximately reproduce the observed
$\afe$-log(age) trend and none of them reproduces the observed $\ohh$-log(age)
trend. The black curve shows a reference model with $\eta=2$,
$\taustar=4\Gyr$, and an exponential SFH with $\tausfh=6\Gyr$. These values
are plausible choices for the solar annulus and yield approximately solar
abundances at late times. Lowering $\eta$ to one (red curve) raises the
equilibrium $\ohh$ (eq.~\ref{eqn:ZoEq}) and shifts the $\afe$ equilibrium
slightly upward. Reducing $\tausfh$ to 4~Gyr (sienna curve) raises the
equilibrium $\ohh$ slightly, and it lowers the $\afe$ track because more
rapidly declining star formation makes the ratio of delayed SNIa to
instantaneous CCSN higher at any given time. Doubling the SFE time-scale to
$\taustar=8\Gyr$ (magenta curve) slows the approach to equilibrium by
increasing $\taudep$, with a small shift in the equilibrium abundance.
Keeping $\taustar=4\Gyr$ while changing from exponential to
linear--exponential SFH has a stronger impact (blue curve) because of the much
slower approach to equilibrium (eq.~\ref{eqn:ZoEvolLinExp}). 

The lower panels of Fig.~\ref{fig:onezone} change the $x$-axis from 
logarithmic to linear.
In this representation, it is more evident that some models
still have gently rising $\ohh$ at $t=10\Gyr$, and that $\afe$ is still
declining at late times in all models because of the long tail of the
$t^{-1.1}$ DTD for SNIa. The dotted black curve shows the effect of changing
to a single-exponential DTD with $\tau_{\rm Ia}=1.5\Gyr$, the form used for
most of the calculations in WAF17 and \cite{andrews2017}. With fewer SNIa at
long delay times, this model settles to an equilibrium $\afe$ by $t \approx
6\Gyr$. The dashed black curve shows the effect of changing the minimum SNIa
delay time from 0.15~Gyr to 0.05~Gyr while retaining the $t^{-1.1}$ form. Here
$\afe$ declines more rapidly at early times but has similar late-time behavior;
in an $\afe$-log(age) plot the impact of this change is small. Changes in the
SNIa yield or DTD have no impact on the $\ohh$-time relation, which is the
reason we have chosen to focus on modeling this relation rather than $\feh$
vs.\ time.

The WAF17 analytic solutions require metallicity-independent yields, but
they can be adapted to elements with different relative contributions
of CCSN and SNIa. To express the behaviour in general terms, it is
useful to define
\begin{equation}
\Delta\ofe = -\log_{10}\left[1+\ZfeIa(t)/\Zfecc(t)\right]~,
\label{eqn:deltaofe}
\end{equation}
the change in $\ofe$ relative to the plateau value at early times,
when CCSN dominate the enrichment for both O and Fe.
Here $\ZfeIa(t)$ and $\Zfecc(t)$ represent the ISM iron mass fraction
contributed by SNIa and CCSN, respectively, at time $t$.
Define
\begin{equation}
F_X = {\mxIa/\mxcc \over \mfeIa/\mfecc}~,
\label{eqn:fx}
\end{equation}
where $\mxIa$ and $\mxcc$ are the IMF-averaged yield of element X
from SNIa and CCSN. For a ``pure-$\alpha$'' element produced
entirely by CCSN, $F_X=0$, and for iron $F_X=1$.
Using Equation~\ref{eqn:deltaofe} 
and equation 109 of WAF17, one can show that
\begin{equation}
\Delta\xfe = \Delta\ofe + \log_{10}
 \left[1+F_X\left(10^{-\Delta\ofe}-1\right)\right]~.
\label{eqn:deltaxfe}
\end{equation}
Equation~\ref{eqn:deltaxfe} can be used to convert a
$\Delta\afe$-age relation into a $\Delta\xfe$-age relation for other elements.

Fig.~\ref{fig:fx} shows $\Delta\xfe$-age predictions for several choices
of $F_X$, using the parameters of the fiducial (black curve) one-zone
model from Fig.~\ref{fig:onezone}. For reference, the yield calculations
of \citet{andrews2017} imply $F_X=0.24$ and 0.36 for Si and S,
respectively, given the values $\mfeIa=\mfecc=0.0015$ that we are using here.
Roughly speaking, the $\Delta\xfe$-age relation for an element
with $F_X=0.25$ (0.50) resembles the $\Delta\afe$-age relation with the
vertical drop scaled by a factor of 0.67 (0.39).
The normalisation of an $\xfe$-age relation further depends on 
the value of $\mxcc/\mfecc$ relative to the solar X/Fe ratio.
The observed [X/M]-age relations for Si, S, Ca, and Ti do not
obviously show the trend predicted by Fig.~\ref{fig:fx} for
steadily increasing $F_X$, and this discussion does not explain
the slight increase of mean age seen for the lowest bins of
Si/M, Ca/M, and Ti/M. The observed [Ni/M]-age trend, on the 
other hand, looks roughly like the model prediction for 
$F_X \approx 0.35$. Again, this is not consistent with theoretical predictions of SN yields and suggests the nucleosynthesis of Ni is not well understood.

\begin{figure*}
\centerline{
\includegraphics[clip, trim=0cm 0cm 0cm 0cm, width=0.6\textwidth]{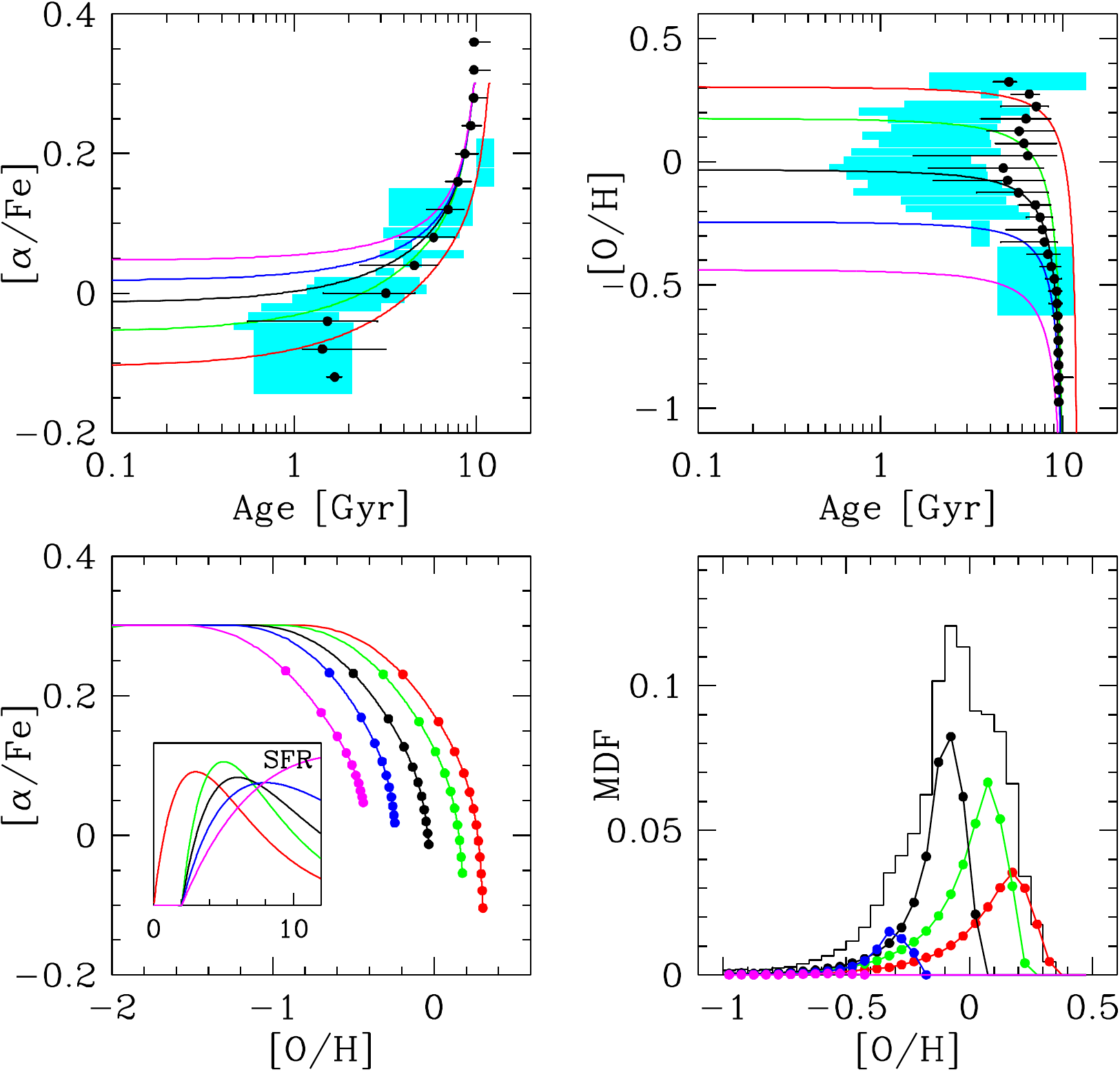}
}
\caption{Mixture model of the age-abundance distributions. In the lower left
panel, curves show the $\afe-\ohh$ evolution of five one-zone models with the
parameters given in Table~\ref{tbl:mixmodel}, with points marking 1~Gyr
intervals and inset panel showing the star formation histories.
In the upper panels, points show the
median age of stars in 0.04~dex bins of $\afe$ (left) or 0.05~dex bins of
$\ohh$ (right), and line segments mark the 16~per cent-84~per cent range of ages in the
bin. Cyan rectangles represent the APOGEE measurements.  The individual
model evolutionary curves are shown in the upper panels for reference.
In the lower right panel, the black histogram shows the $\ohh$ MDF of the
solar annulus at $t=12\Gyr$ and colored lines show the contributions
of individual annuli to this MDF.}
\label{fig:mixmodel}
\end{figure*}

All of these simple models predict one-to-one age-abundance relations and a
monotonically increasing $\ohh$. To better reproduce the observed
distributions, we consider a more realistic model in which inflow, outflow,
SFE, and SFH parameters vary with Galactocentric radius and radial migration
mixes stars born at smaller and larger radii into the solar neighbourhood. One
can view this mixture model as a simplified and parameterized representation of
the behavior in more sophisticated models such as \cite{Schonrich2009} and
\cite{Minchev2013}. Our choice of parameters for the five Galactic annuli is
given in Table~\ref{tbl:mixmodel}, and the $\afe-\ohh$ evolutionary tracks of
these models appear in the lower left panel of Fig.~\ref{fig:mixmodel}. We
have chosen linear--exponential SFHs for these models both because we consider
them more realistic than exponentials and because the slower approach to
equilibrium helps reproduce the observed behavior. The inner annulus evolves
from $t=0$ to $t=12\Gyr$, while the other models start at $t=2\Gyr$. An
earlier commencement of star formation in the inner Galaxy is physically
reasonable, and we find it difficult to reproduce the observed ages of the most
metal-rich stars without giving the inner annulus a head start. From the inner
to outer Galaxy, we increase $\eta$ to produce a negative metallicity gradient,
decrease the star formation efficiency to represent the impact of lower gas
surface density, and increase $\tausfh$ to produce more extended star
formation.

\begin{figure*}
\centerline{\includegraphics[width=5.5truein]{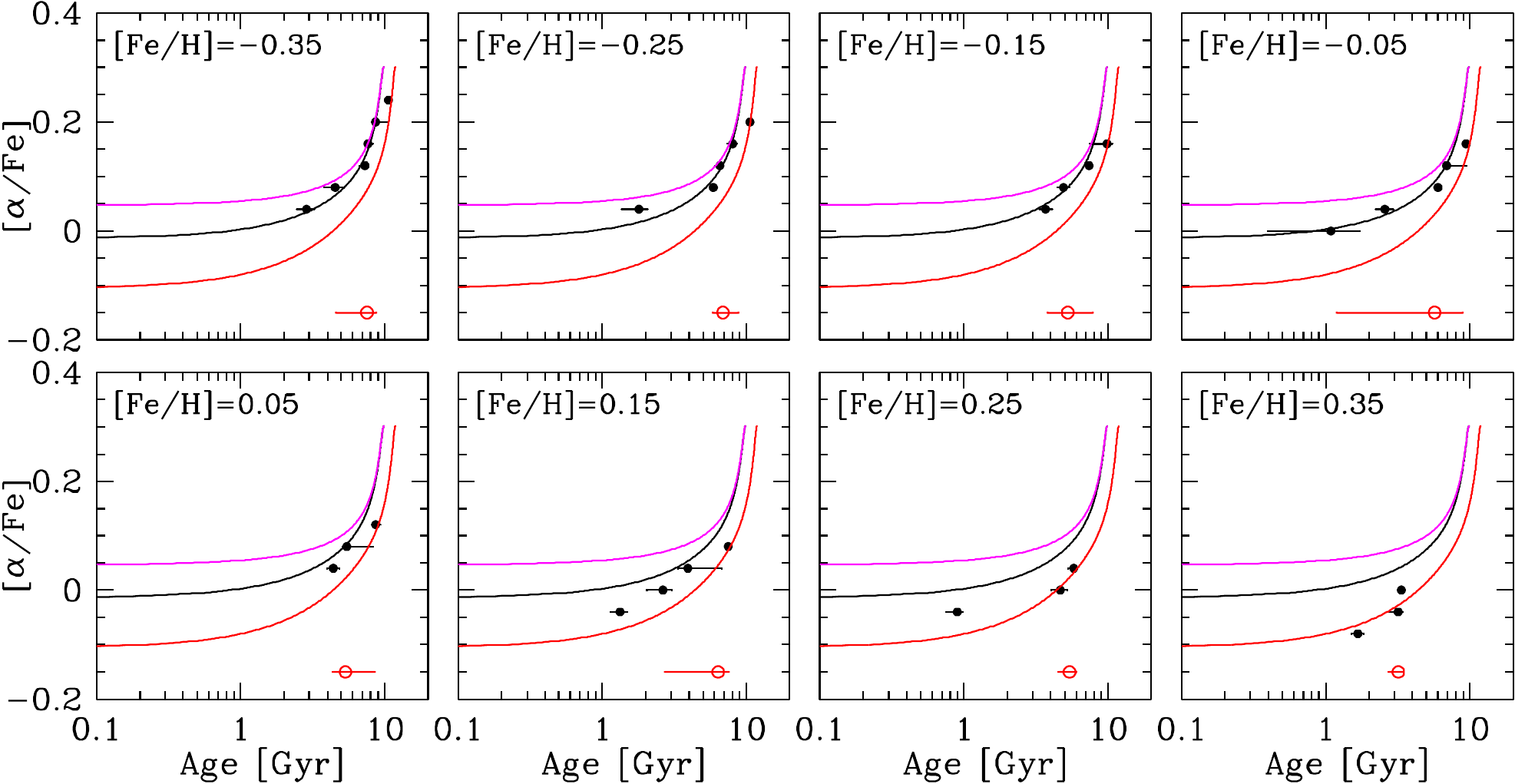}}
\caption{$\afe$-age relation predicted by the mixture model in
0.1~dex bins of $\feh$, for comparison to Fig.~\ref{fig:age_alpha_fe}.
In each panel, filled circles and horizontal bars show the median and 16--84~per cent
range of age in 0.04~dex bins of $\afe$, while the open circle and horizontal
bar represent the median age and 16--84~per cent range for the full $\feh$ bin.
Curves show the central model and the inner and outermost models
contributing to the mixture, for reference. In contrast to 
Fig.~\ref{fig:mixmodel}, we do not include boxcar smoothing of
the abundance distributions prior to computing age ranges.
} 
\label{fig:mixmodel_alpha}
\end{figure*}

\begin{table} 
\centering 
\begin{tabular}{ccccc} 
\hline
$R$ (kpc) & $\eta$ & $\taustar$ & $\tausfh$ & $t_{\rm start}$ \\
\hline 
3-5 & 1.0 & 1.5 & 3.0 & 0 Gyr  \\ 
5-7 & 1.5 & 2.0 & 3.0 & 2 Gyr  \\ 
7-9 & 2.5 & 3.0 & 4.0 & 2 Gyr  \\ 
9-11 & 4.0 & 4.0 & 6.0 & 2 Gyr  \\ 
11-13 & 6.0 & 8.0 & 12.0 & 2 Gyr \\ 
\hline
\end{tabular} 
\caption{Mixture Model Parameters: All models adopt a linear--exponential SFH, IMF-averaged yields $\mocc=0.014$, $\mfecc=\mfeIa=0.0015$, and a
two-exponential DTD for SNIa that approximates a $t^{-1.1}$ power-law with minimum delay time of 0.15~Gyr (see WAF17). Units of $\taustar$, $\tausfh$, and $t_{\rm start}$ are~Gyr. } 
\label{tbl:mixmodel} 
\end{table}

\begin{figure*} 
\centerline{\includegraphics[width=5.5truein]{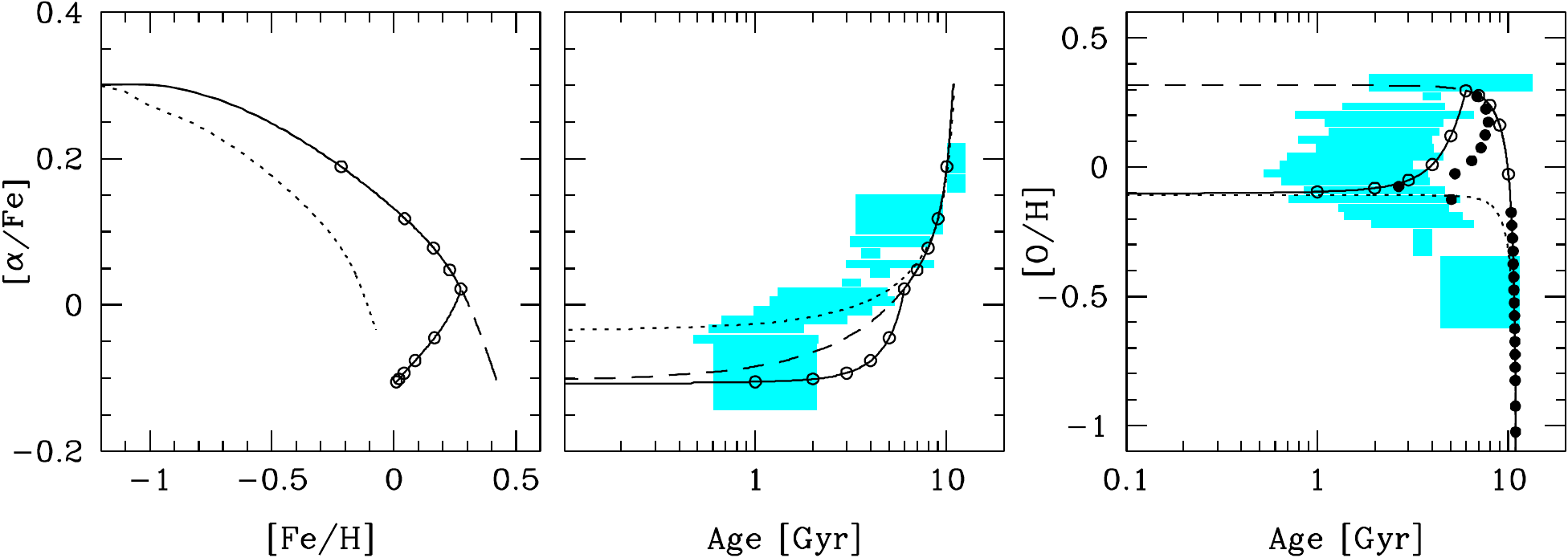}}
\caption{A one-zone model with a discontinuous change of parameters, from
$(\eta,\taustar,\tausfh)=(1.0,2.0,5.0)$ for $t \leq 5\Gyr$ to
$(\eta,\taustar,\tausfh)=(3.0,4.0,10.0)$ for $5 < t \leq 11\Gyr$. In each
panel, the solid curve shows the model evolutionary track with open circles at
$1\Gyr$ intervals. In the right panel, filled circles show the mean age of
stars in 0.05~dex bins of $\ohh$. Cyan rectangles in the middle and right
panels show the APOGEE measurements. In all panels, dashed and dotted curves
show the tracks predicted by the first and second models, respectively, if they
are run over 11~Gyr without parameter changes. } 
\label{fig:reverse}
\end{figure*}

The expected level of radial migration is difficult to predict from first
principles because it is sensitive to properties such as disc velocity
dispersion and gas fraction, bar strength, and satellite perturbations (see,
e.g., \citealt{Roskar2008}; \citealt*{Bird2012}; \citealt{Minchev2013}), whose Galactic history is
largely unknown. Here we adopt a simplified model of radial migration in which
the stars formed in annulus $i$ at time $\tform$ are distributed in
radius radius at the final epoch $t_0=12\Gyr$ with a Gaussian distribution
$p(R) = (2\pi\sigma^2)^{-1/2}\exp\left[-(R-R_i)^2/2\sigma^2\right]$ where
$R_i$ is the central radius of the annulus and the dispersion is
\begin{equation}
\sigma = \sigma_0 \left(1-\tform/t_0\right)^{1/2}~,
\label{eqn:sigmat}
\end{equation}
growing $\propto (\Delta t)^{1/2}$ as expected for a diffusion process.
We adopt $\sigma_0=2.5\kpc$ for all five annuli.
We also account for the exponential surface density profile of the
disk by normalizing the star formation history of each annulus
to produce a time-integrated stellar mass proportional to 
$e^{-R_i/R_H}$ with $R_h=2.5\kpc$.  Thus, the number of stars of
age $t-\tform$ at the solar radius $\Rsol=8\kpc$ from annulus $i$ is
proportional to 
\begin{equation}
F_{\rm sol}(R_i,\tform) = {\rm SFR}(R_i,\tform) e^{-(R_i-\Rsol)/R_h}
  e^{-(R_i-\Rsol)^2/2\sigma^2~}.
\label{eqn:FofR}
\end{equation}
This model crudely reproduces the kind of pattern seen in the disk
galaxy simulations of Bird et al.\ (\citeyear{Bird2013}, fig.~15)
or Minchev et al.\ (\citeyear{Minchev2013}, fig.~3), in which young
stars in the solar neighbourhood formed near $\Rsol$ but older stars
come largely from the inner galaxy.
The lower right panel of Fig.~\ref{fig:mixmodel} shows the histogram
of $\ohh$ at the solar annulus at $t=12\Gyr$, and the contributions
from each annulus to this MDF.  The metal-rich tail is dominated 
by stars from the $R_i=4\kpc$ annulus, while the core of the
MDF is dominated by stars from the $R_i=6\kpc$ and $R_i=8\kpc$ annuli.
The contribution of the $R_i=10\kpc$ annulus is small because of
the exponential surface density factor, and that of the $R_i=12\kpc$
annulus is negligible because of both the surface density factor
and the small fraction of stars that migrate 4 kpc.

The principal results of the mixture model are shown by the points and range
bars in the upper panels of Fig.~\ref{fig:mixmodel}, which mark the median
age and the 16--84~per cent range of ages from the population mixture computed in
bins of 0.04~dex in $\afe$ or 0.05~dex in $\ohh$. After assigning stars to
bins, we apply a 3-cell boxcar smoothing along the $\afe$ and $\ohh$ dimensions
before computing these statistics. This convolution mitigates the artificial
discreteness of our five-zone mixture, which would in practice be smoothed by
observational errors, by continuous gradients, and by scatter about idealised
evolutionary tracks. The mixture model is reasonably successful in reproducing
the observed $\afe$-log(age) trend and the scatter in age at a fixed $\afe$.
It does not reproduce the large age spread at $\afe \approx +0.12$. The
discrepancy in these bins could indicate that some young $\alpha$-enhanced
stars are produced by a mechanism not represented in this model, or that some
of the apparently young stars are ``rejuvenated'' products of binary stellar mergers
\citep{tayar2015, Jofre2016}, or that other age-determination errors have artificially
broadened the age range.

The mixture model reproduces the observed $\ohh$-log(age) trend better
than an individual one-zone model, but the match is still rather poor.
From the individual zone evolutionary tracks and MDF contributions
one can see the basic structure of the model.  For $\ohh<0$ the population
is dominated by stars formed in the solar annulus, and the age trend
follows that of the $R_i=8\kpc$ model.  Just above $\ohh=0$ the
dominant population shifts to stars formed in the $R_i=6\kpc$ annulus,
which are older at a given $\ohh$.  A similar transition to the $R_i=4\kpc$
annulus, and another, smaller jump of median age, occurs at $\ohh \approx +0.2$.
The model
successfully produces a large age spread near solar metallicity and an increase
of median age at super-solar metallicity. However, despite the 
model's adjustable parameters,
we have not found a mixture that produces a median age
as young as 2~Gyr at solar metallicity or the smooth increase of age with
increasing $\ohh$ at super-solar metallicity. 
The model as implemented here does not take
account of the vertical distribution of stars, in particular the observed
trend that older stellar populations have larger vertical scale heights.
Incorporating this effect together with the local neighbourhood selection
of our observational sample might lead to a younger predicted age at solar
metallicity, since the older stars of the {\it in situ} zone would be
less likely to be included in the sample.

Fig.~\ref{fig:mixmodel_alpha} shows the predicted $\afe$-age relation
in 0.1~dex bins of $\feh$, similar to Fig.~\ref{fig:age_alpha_fe}.
For simplicity of interpretation, we have omitted the boxcar smoothing
of abundance distributions used for the predictions in 
Fig.~\ref{fig:mixmodel}. 
In agreement with the observations, each individual $\feh$ bin shows
an $\afe$-age relation similar to the global relation, and from
inspection of Fig.~\ref{fig:mixmodel} it is easy to see why.
First, the $\afe$-age relations of the individual models composing
the mixture are similar to begin with, differing mainly in the 
asymptotic value at low age. Second, in any given $\feh$ bin,
the populations of one or two zones dominate, and the $\afe$-age
relation for the bin tracks the one-to-one relation of the
dominant zones. The mixture model predicts very small scatter
in the $\afe$-age relation within $\feh$ bins. The observations
show substantially larger age dispersion, implying an additional
source of scatter in the age-abundance relations beyond simple
mixing of individually smooth one-zone models.

WAF17 note that even a one-zone model can exhibit non-monotonic metallicity
evolution if the outflow mass-loading increases at late times (e.g., as a
consequence of decreasing gas surface density). Fig.~\ref{fig:reverse} shows
an example of such a model using the 2-phase analytic solution described in \S
4.2 of WAF17. For $t \leq 5\Gyr$ the model has an exponential SFH with
$\eta=1.0$, $\taustar=2\Gyr$, $\tausfh=5\Gyr$ and evolves to $\ohh \approx
+0.3$, $\afe \approx 0$. At $t=5\Gyr$ the model parameters change
discontinuously to $\eta=3.0$, $\taustar=4\Gyr$, $\tausfh=10\Gyr$, with a lower
equilibrium $\ohh$. The continuing downward evolution of $\afe$ is driven by
SNIa from stars that form before 5~Gyr but explode and deposit their iron after
5~Gyr. While the evolution of $\afe$ is monotonic, the age-$\ohh$ relation is
double valued for $\ohh > -0.1$. Filled circles in the right panel show the
mass-weighted mean stellar age. Clearly this model is not a good fit to the
data, but it illustrates an alternative mechanism for producing young mean ages
near solar metallicity and a smooth increase of age at higher metallicity. An
outflow mass-loading that grows at late times is physically plausible, and a
combination of this mechanism with radial migration might be important for
reproducing the full structure of the observed age-abundance relation.

Part of the difficulty in reproducing the young age at solar metallicity
in the mixture model of Fig.~\ref{fig:mixmodel} is that each individual
zone approaches equilibrium and produces stars at nearly the same
metallicity over several Gyr.  This property is evident in the nearly
vertical evolutionary tracks in the lower left panel and the nearly
horizontal $\ohh$-log(age) relations in the upper right panel.
One can mitigate this effect by allowing $\eta$ to slowly decrease
in time, so that the equilibrium abundance itself grows steadily
with time.  We have experimented with this type of model, adapting
the procedure described by equations 115 and 116 of WAF17 to construct
analytic solutions in which $(1+\eta-r)$ decreases linearly in each
annulus over $t=0-12\Gyr$.  With continuously increasing $\ohh$
in individual zones we are able to get a slightly better match to
the observed $\ohh$-log(age) distribution, but in our limited
explorations we still have not found a model that produces a median
age below $4\Gyr$ near solar metallicity.  These models still have
smooth star formation histories, and it could be that star formation
bursts or the vertical selection effect mentioned above are
crucial for reproducing these APOGEE results.

There are several directions for improving this kind of multi-zone modeling.
One is to carry out more systematic sampling of the high-dimensional parameter
space, e.g.\ via Markov Chain Monte Carlo, which requires first deciding on a
figure of merit for evaluating agreement with observations. This approach
might involve forward-modeling the impact of observational errors instead of
Bayesian inference of age-distribution parameters as an intermediate step. A
second direction is to use numerical simulation predictions as a source of
informed priors for radial migration prescriptions and parameter ranges. A
third direction, probably essential for breaking parameter degeneracies, is to
expand the set of observational constraints, including MDFs and $\afe-\feh$
tracks as a function of Galactic position \citep{Hayden2015}, present-day ISM
abundance gradients, and age-abundance distributions over a wider range of the
disc. Robust inferences must account for uncertainties in IMF-averaged yields
and SNIa DTDs, and improving external constraints on these quantities would
mitigate degeneracies with chemical evolution parameters that describe
outflows, star formation efficiency, and star formation histories. We
anticipate progress along all of these directions in future work. The
difficulty of finding a fully successful fit to the data even with the many
adjustable parameters of our mixture model implies that any {\it ab initio}
model that reproduces the measured age-abundance distributions with few or no
adjustments has scored a significant success.

\subsection{Chempy model} 
\label{sec:Rybizki}

\begin{figure*}
\centering 
\includegraphics[clip, angle=90, trim=0cm 0.8cm 0.9cm 0.8cm, width=0.8\textwidth]{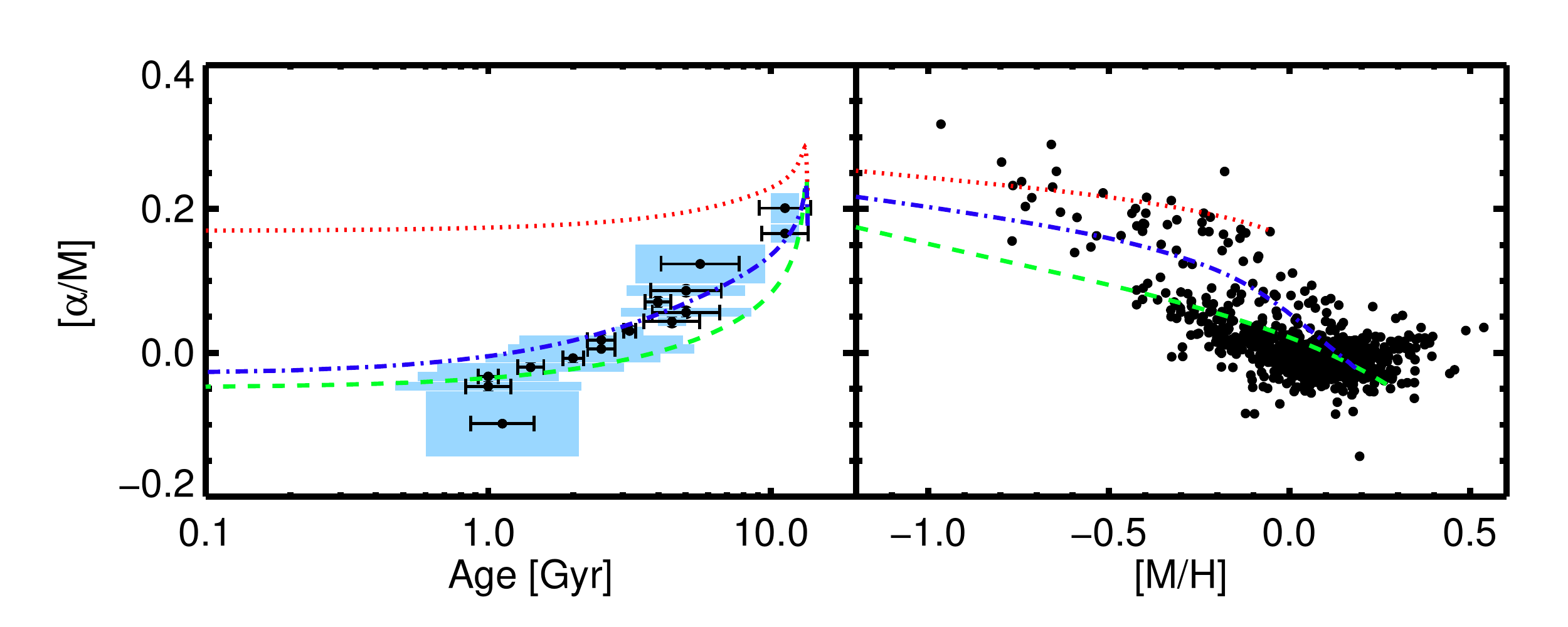} 
\includegraphics[clip, angle=90, trim=0.5cm 0.8cm 0.9cm 0.8cm, width=0.8\textwidth]{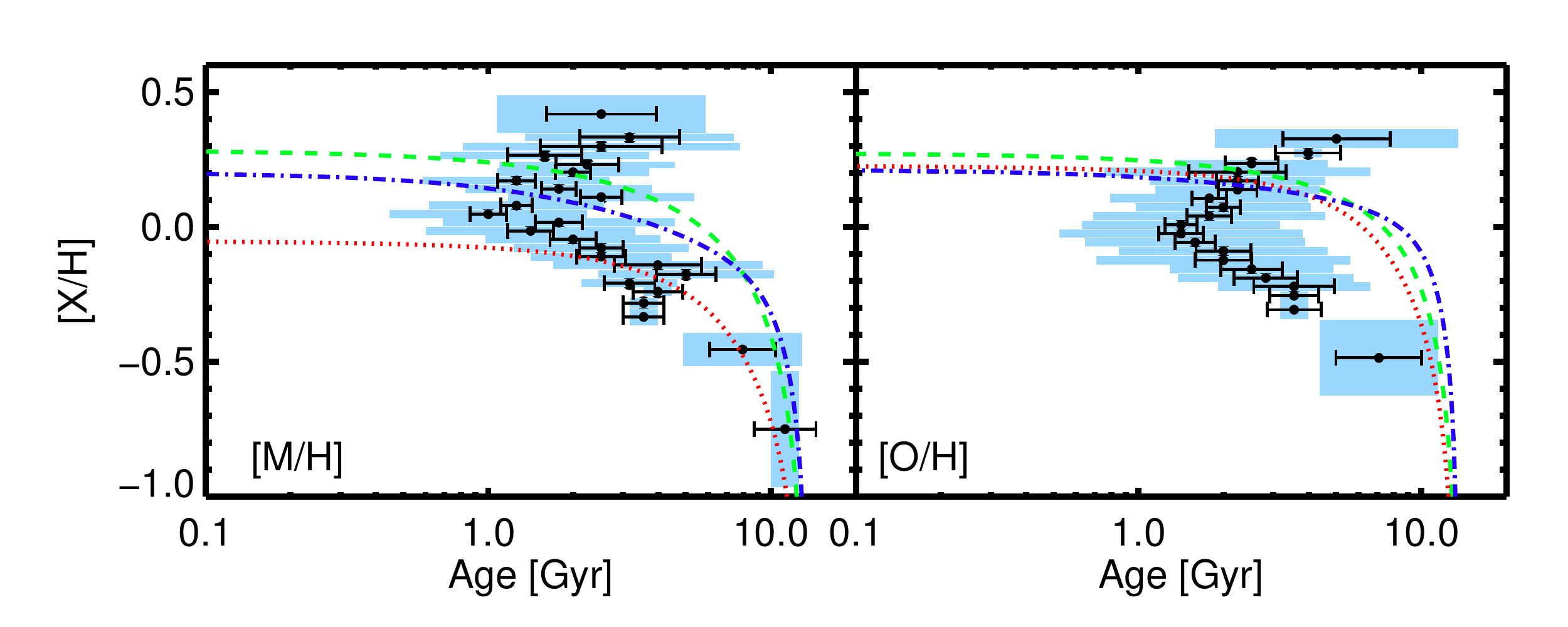} 
\caption{ The upper
left panel and both bottom panels are the same as Fig.~\ref{fig:age_al}
and~\ref{fig:age_fe_o}, but showing the observed age-abundance relation for [\al/M],
[M/H], and [O/H], as well as [\al/M] vs [M/H] with the R17 GCE models shown.
The lines are the results of the R17 GCE model using the default yields fit to
the Sun (dashed green) and Arcturus (dotted red), and the alternative yield set
fit to the Sun (dash-dotted blue).} 
\label{fig:age_MH_CEM} 
\end{figure*}

{\it Chempy} \citep*[][hereafter R17]{Rybizki2017} is a flexible, open-source
code developed for modeling Galactic chemical evolution. 
It incorporates detailed predictions for the yields of many chemical elements
based on the prescriptions of \citet{Karakas2010}, \citet*{Nomoto2013}, and
\citet{Seitenzahl2013} for asymptotic giant branch (AGB), core-collapse supernovae,
and Type Ia supernovae feedback respectively (the ``default'' yield set).
An ``alternative'' yield set instead uses the
prescriptions of \citet{Ventura2013}, \citet{Chieffi2004},and
\citet{Thielemann2003} for these three enrichment processes.
The code functions in a Bayesian framework to constrain physical 
parameters that describe the star formation efficiency, the star 
formation history, the outflow efficiency, and the initial mass
of the gaseous corona. It simultaneously constrains parameters
that describe the stellar and supernova population, namely the
high mass slope of the IMF, the minimum delay time for SNIa, and
the normalisation of the SNIa DTD (assumed to have a $t^{-1.1}$ form).
R17 infer optimal parameter values by fitting the detailed chemical
abundance pattern of the Sun, or Arcturus, or present-day B-stars, 
together with the observed CCSN/SNIa ratio from \citet{Mannucci2005}
and the coronal metallicity measured by \citet{Fox2016}.
Here we compare to their ``Sun+'' model computed with both the default
and alternative yield sets and their ``Arcturus+'' models computed with the 
default yield set (see Table 4 or R17 for parameter values).
These are one-zone models; R17 also construct a multi-zone model
in which the solar, Arcturus, and B-star abundances are fit 
simultaneously with the same stellar population parameters but with
independent star formation and outflow parameters to represent
different birth zones. 

In the terminology of the previous section, the parameters of the
Sun+ model correspond approximately to $\taustar=2\Gyr$, $\eta=1$,
and a linear--exponential SFH with $\tausfh=3\Gyr$.
However, while the WAF17 models assume primordial composition infall,
the R17 models have metal exchange between the ISM and the corona.
More importantly, the R17 models incorporate nucleosynthesis 
predictions for many elements, including metallicity dependent yields.
Although R17 predict significant AGB contributions to C and N,
we do not include those in our comparison here because the models
do not incorporate internal mixing on the giant branch. The predicted AGB
contributions to the other elements considered here are below 10~per cent,
far below in most cases (see Fig.~13 of R17).

Fig.~\ref{fig:age_MH_CEM} shows the [M/H]-age relation from Fig.~\ref{fig:age_fe_o}
with the R17 GCE model predictions overplotted as lines. 
The Sun+ model with the default yield set gives the best match to the
[\al/M]-age relation, and its evolutionary track in the [\al/M]-[M/H]
plane is similar to the observed low-[\al/M] sequence.
The predicted drop in [\al/M] is faster at early times than
the observations imply, which suggests that SNIa iron enrichment
is becoming important too quickly.
The Arcturus-normalised models predict only a small drop in [\al/M]
with time, which is due to its lower fractional SNIa contribution
to iron-peak elements. 
They yield a poor match to the observations, but qualitatively the behavior is 
similar to the one-zone WAF17 models. In particular, the Arcturus model is very 
similar to the inner-most WAF17 model.
The predicted tracks of [M/H]-age and [O/H]-age resemble those of
the one-zone models in Fig.~\ref{fig:onezone}, and like those models
they do not reproduce
the observed increase of mean age above solar metallicity.
All three models approach solar [O/H] too quickly, and the
Sun+ model also rises too quickly in [M/H].

\begin{figure*}
\centering
\includegraphics[clip, angle=90, trim=0.5cm 0.8cm 0.9cm 0.8cm, width=1.0\textwidth]{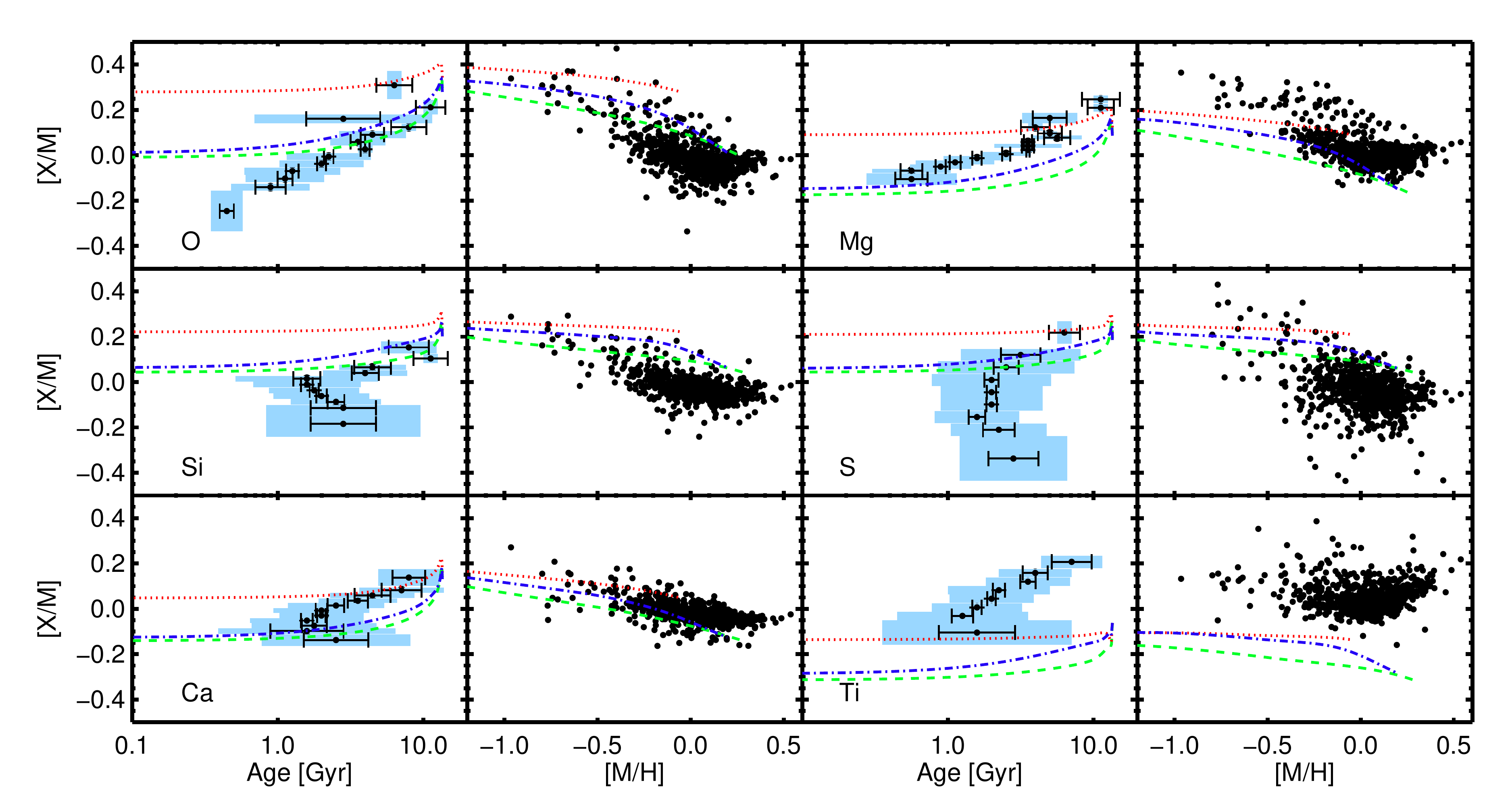} 
\caption{Same as
Fig.~\ref{fig:age_OMgSi} for all \al-elements comparing to GCE models. The lines
are the results of the R17 GCE model using the default yields fit to the Sun
(dashed green) and Arcturus (dotted red), and the alternative yield set fit to
the Sun (dash-dotted blue).} 
\label{fig:age_alpha_CEM} 
\end{figure*}

\begin{figure*}
\centering 
\includegraphics[clip, angle=90, trim=0.5cm 0.8cm 0.9cm 0.8cm, width=1.0\textwidth]{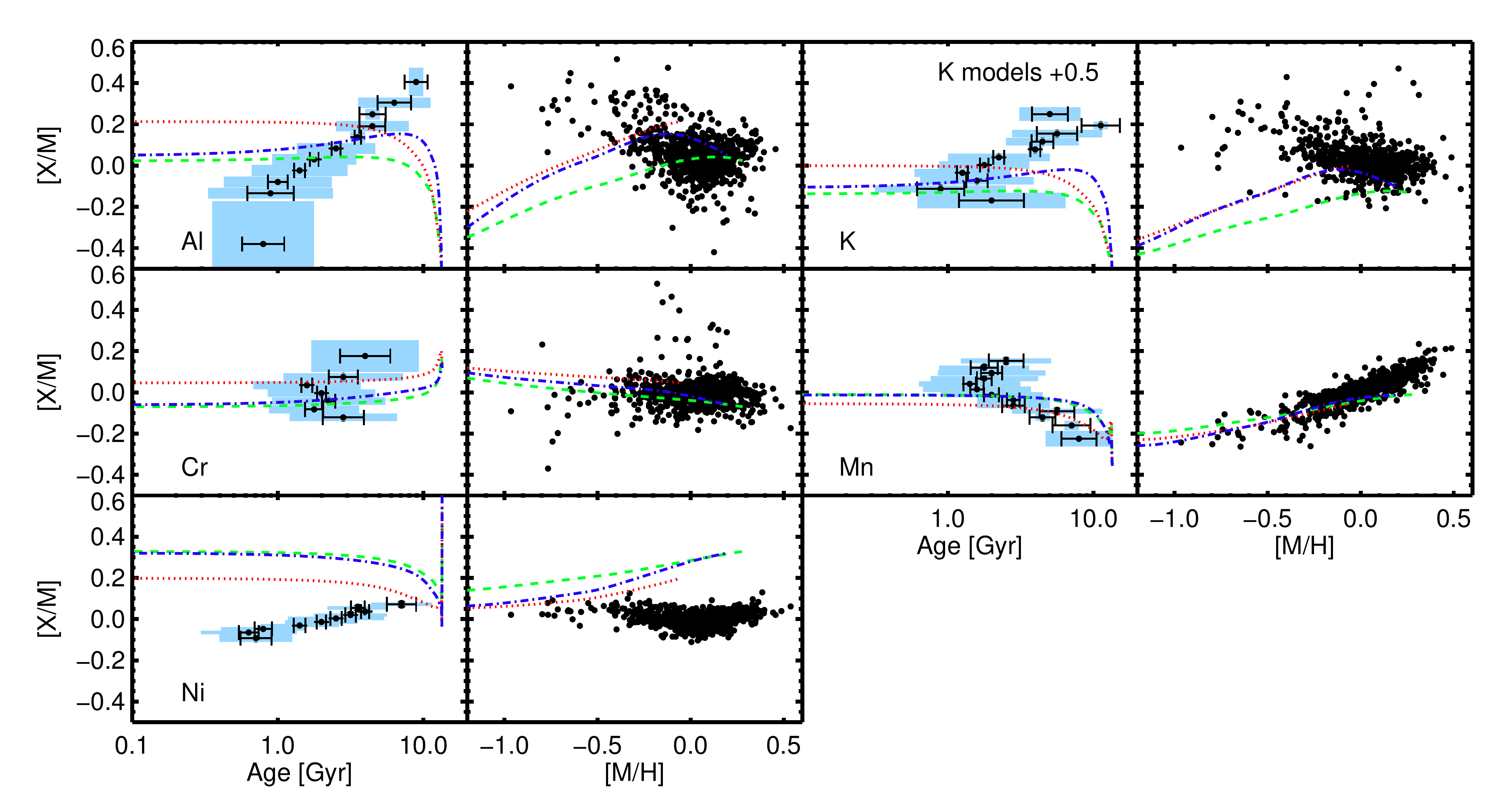} 
\caption{Same as
Fig.~\ref{fig:age_OMgSi} for the light odd-Z and Fe-peak elements comparing to GCE models.
The lines are the results of the R17 GCE model using the default yields fit to
the Sun (dashed green) and Arcturus (dotted red), and the alternative yield set
fit to the Sun (dash-dotted blue).} 
\label{fig:age_oddfe_CEM} 
\end{figure*}

Fig.~\ref{fig:age_alpha_CEM} and~\ref{fig:age_oddfe_CEM} show the comparisons
to the R17 GCE models for the individual elements. These figures are the same
as Fig.~\ref{fig:age_OMgSi} through~\ref{fig:age_CrMnNi}, with the three {\it Chempy}
models overplotted for both the age and [X/M] vs [M/H] distributions. 
Rather than discuss individual element comparisons, we make a few
general observations, focusing on the Sun+ model. For the $\alpha$
elements, the predicted [X/M]-age tracks differ in the normalisation
at early times, which reflects the CCSN yield ratios relative to solar
values, and in the drop between this initial ratio and the late-time
ratio, which mainly reflects the relative contribution of SNIa to
this element. In particular, the predicted SNIa contribution 
is negligible for O and Mg but significant for Si, S, Ca, and Ti
(see Fig.~13 of R17), so these elements show a smaller predicted
drop. However, the element-by-element variations predicted by
{\it Chempy} do not produce good agreement with the APOGEE data.
The observed [X/M]-age relations for the $\alpha$ elements are
fairly similar for the most part, and to the extent that they 
differ it is not in the ways predicted by the models.

For light odd-Z elements, the R17 models predict a rapid rise in
[X/M] at early times because of metallicity-dependent yields,
followed by a plateau in [X/M] at late times. However, the models predict very few stars 
at the immediate start of star formation, therefore the sharp rise and peak would not be 
observable in the stellar population. The R17 models are not consistent with the 
[X/M]-[M/H] distributions of the data. The observed age trends for Al and K qualitatively
resemble the observed (and predicted) trends for $\alpha$ elements
rather than the flat predictions for odd-Z elements. This disagreement
suggests that the yield models predict overly strong metallicity dependence,
and that CCSN or some other nucleosynthetic source produce
light odd-Z elements even when the metallicity of the stellar population is low.

For Mn and Ni the model predictions again rise
at early times because of metallicity-dependent yields. 
The predicted [Mn/M]-age and [Mn/M]-[M/H] trends are in strikingly
good agreement with the data, the best match found in these figures.
As mentioned in Section~\ref{sec:fepeak_disc}, NLTE corrections could
be important for [Mn/M], in which case this good agreement would
be a coincidence.
As mentioned in \S \ref{sec:fepeak_disc}, the theoretical Ni trend does not agree with the observed trend.
The models predict a shallow trend of [Ni/M] with age that is
opposite in direction to the observed shallow trend.
This discrepancy could reflect incorrect metallicity-dependence
of Ni yields in the models, or a CCSN contribution to Ni that is
too small, or both.

R17 find that the observed solar abundance pattern, CCSN/SNIa ratio, and
corona metallicity are sufficient to provide interestingly tight 
constraints on the seven parameters of their one-zone chemical
evolution model. The poor agreement with the observed abundance-age
relations found here indicates that these models are still missing
one or more physical ingredients that have a major impact on these
relations. The analysis of \S\ref{sec:WAF} suggests that radial
mixing of stellar populations is at least one of those ingredients.
However, the relative behaviour of different elements depends
largely on yields, and the disagreements in
Fig.~\ref{fig:age_alpha_CEM}-\ref{fig:age_fepeak_CEM} probably indicate
that even the most sophisticated current nucleosynthesis calculations
still have substantial errors for some elements.

\section{Conclusions} 
\label{sec:conclusion}

As described in F16, ages of red giant stars can be determined through Bayesian
isochrone matching with 0.16~dex uncertainty if the distances are known
precisely. However, these age uncertainties, as well as those possible for CN ages ($\sim0.20$~dex), 
are still high for examining detailed age trends across the Galactic disc using individual stars. 
In this work, we use the hierarchical modelling technique described in
F16 to model the SFH of our sample of local red giants as a function of
individual elemental abundances and present age-abundance trends for the solar
neighbourhood. Using APOGEE DR14 parameters, we confirm the steeply increasing
[\al/M]-age trend found in F16, and also present age trends for the overall
metallicity and 17 individual elements. We do note that our analysis does not
account for any possible systematic uncertainties due to the spectroscopic
methods or the stellar models.

Using both [M/H] and [O/H], we find that the most metal-rich stars in our
sample are older than the solar metallicity stars. This behavior is consistent
with the [\al/M]-age relation and the [\al/M] vs [M/H] distribution of the sample.
The solar metallicity bins contain stars with a large range of [\al/M]
abundances, including the lowest, sub-solar [\al/M] stars, while the highest
metallicity bins have a smaller range in [\al/M] abundances centred around solar.
Through modeling the SFH of stars binned in both [M/H] and [\al/M] we find that
the metal-rich stars have a gradient in age with [\al/M] at each [M/H]. At any
given [M/H] the high-[\al/M] stars are older than the low-[\al/M] stars.
However, the youngest stars are still those with solar metallicity and
sub-solar [\al/M]. As these are the youngest stars, they are the most likely to 
have formed at the solar radius. Because the metal-rich, solar [\al/M] stars are older, they probably 
did not form in the solar neighbourhood. They could have radially
migrated from elsewhere, perhaps the inner disc, where star formation occurred
more rapidly and perhaps started earlier, resulting in high metal content of
the ISM at earlier times.

We confirm the strong age-abundance trends of C and N in giant stars due to
internal mixing discussed in work by \citet{Masseron2015} and
\citet{Martig2016}. Our [C/N]-age relation using ages derived independently of the 
C and N abundance suggests the minimum uncertainty in the age derived using 
the CN abundance technique is approximately 0.2~dex in log(age).
The age trends of the individual \al-elements are
consistent with the overall [\al/M] trend, and are especially tight and steep
in O and Mg. Si and Ca deviate from the [\al/M] trend in the lowest abundance
ratio bins, and S has a very broad distribution with only a weak age trend in
the highest and lowest abundance ratio bins. Al and K have \al-like age
trends, while the Na and P trends are consistent with flat or weakly decreasing
in age with abundance ratio. V, Cr, and Co also have flat age-abundance ratio
trends. Mn has an age trend that is very similar to [M/H] and Ni has a very
tight, \al-like behavior.

These observations are crucial for providing strong constraints on the
parameters of GCE models. As we show with the comparisons to WAF17 analytic GCE
models and the flexible `leaky box' GCE model {\it Chempy} from R17, single
zone models cannot reproduce the full age-abundance sequences. Using the WAF17
models, a mixture model of five single zones can reproduce the continuous
age-[\al/M] trend but not large age dispersion in the high [\al/M] stars. The
mixture model is able to produce super-solar metallicity stars with older ages
than the mean age of the solar metallicity stars. However, the mean age of the
solar metallicity stars is not as young as the observations and the mean age
does not increase smoothly with increasing metallicity. The Sun+ model of R17
is consistent with the shape of most of the \al-elements for the higher abundances,
but flattens out at younger ages. The observations, for most elements,
maintain a constant slope or turn over and increase in age at lower abundance
ratios. This is particularly striking in the case of [M/H] and [Mn/M] for
which the {\it Chempy} models are consistent with the lower abundance bins, but
flatten while the observations turn over and increase with age at higher
abundances. These comparisons add evidence in support of the picture 
that individual abundances of stars currently in the
solar neighbourhood cannot be modelled as a single evolutionary zone, but
actually some of these stars were born in different environments and over
time migrated to the solar neighbourhood. In this sense the multi-zone analytic
models are a bit more instructive in constraining the parameters of chemical
evolution elsewhere in the disc. These comparisons also highlight the potential of 
using variations in the abundance-age trends of different elements, even among the 
same element-groups, to help disentangle the effects of Galactic evolution. 

Although observations of the solar neighbourhood are continuing to find an
interesting and diverse population of local stars, such detailed observations
are needed in farther regions of the Galactic disc to form a more complete
picture of the evolutionary history of the Milky Way. Giant stars are
excellent tracers of the disc populations and can be observed to farther
distances than many other stellar types. Using statistical methods such as
those presented in F16, age trends are now possible for large samples of giant
stars with precise atmospheric parameters and distances. Such observations
will be available for a large sample of the Milky Way disc from APOGEE and Gaia
data.

\section*{Acknowledgements}

We thank the anonymous referee for their helpful comments and suggestions. D.K.F. acknowledges funds from the Alexander von Humboldt Foundation in the framework of the Sofja Kovalevskaja Award endowed by the Federal Ministry of Education and Research. J.B. acknowledges the support of the Natural Sciences and Engineering Research Council of Canada (NSERC), funding reference number RGPIN-2015-05235, and from an Alfred P. Sloan Fellowship. DAGH and OZ acknowledge support provided by the Spanish Ministry of Economy and Competitiveness (MINECO) under grant AYA--2014--58082--P.

Funding for the Sloan Digital Sky Survey IV has been provided by the Alfred P.
Sloan Foundation, the U.S. Department of Energy Office of Science, and the
Participating Institutions. SDSS-IV acknowledges support and resources from
the Center for High-Performance Computing at the University of Utah. The SDSS
web site is www.sdss.org.

SDSS-IV is managed by the Astrophysical Research Consortium for the
Participating Institutions of the SDSS Collaboration including the Brazilian
Participation Group, the Carnegie Institution for Science, Carnegie Mellon
University, the Chilean Participation Group, the French Participation Group,
Harvard-Smithsonian Center for Astrophysics, Instituto de Astrof\'isica de
Canarias, The Johns Hopkins University, Kavli Institute for the Physics and
Mathematics of the Universe (IPMU) / University of Tokyo, Lawrence Berkeley
National Laboratory, Leibniz Institut f\"ur Astrophysik Potsdam (AIP),
Max-Planck-Institut f\"ur Astronomie (MPIA Heidelberg), Max-Planck-Institut
f\"ur Astrophysik (MPA Garching), Max-Planck-Institut f\"ur Extraterrestrische
Physik (MPE), National Astronomical Observatories of China, New Mexico State
University, New York University, University of Notre Dame, Observat\'ario
Nacional / MCTI, The Ohio State University, Pennsylvania State University,
Shanghai Astronomical Observatory, United Kingdom Participation Group,
Universidad Nacional Aut\'onoma de M\'exico, University of Arizona, University
of Colorado Boulder, University of Oxford, University of Portsmouth, University
of Utah, University of Virginia, University of Washington, University of
Wisconsin, Vanderbilt University, and Yale University. Collaboration Overview
Start Guide Affiliate Institutions Key People in SDSS Collaboration Council
Committee on Inclusiveness Architects Survey Science Teams and Working Groups
Publication Policy How to Cite SDSS External Collaborator Policy

\bibliographystyle{mnras}
\bibliography{additional,all}

\bsp
\label{lastpage}
\end{document}